\documentclass[aps,prl,twocolumn,showpacs,amsmath,amssymb,superscriptaddress,longbibliography,floatfix]{revtex4-2}

\usepackage[pdftex]{graphicx}
\usepackage[breaklinks,colorlinks=true,citecolor=blue]{hyperref}
\usepackage{braket}
\usepackage{lipsum}
\usepackage{xcolor}
\usepackage{siunitx}

\usepackage{comment}  

\usepackage[capitalize]{cleveref}

\usepackage{glossaries}
\glsdisablehyper
\newacronym{abs}{ABS}{Andreev bound state}
\newacronym{fabs}{FABS}{Fibonacci-Andreev bound state}
\newacronym{jj}{JJ}{Josephson junction}
\newacronym{je}{JE}{Josephson effect}
\newacronym{cpr}{CPR}{current-phase relation}

\definecolor{crimson}{RGB}{255,102,255}
\definecolor{mahogany}{RGB}{192,64,0}
\definecolor{kellygreen}{RGB}{76,187,23}

\newcommand{\md}{\mathrm{d}}

\newcommand{\me}{\mathrm{e}}

\newcommand{\Tr}{\textrm{Tr}}

\usepackage{sidecap,tikz}
\definecolor{lime}{HTML}{A6CE39}
\DeclareRobustCommand{\orcidicon}{\hspace{-1mm}
	\begin{tikzpicture}
		\draw[lime, fill=lime] (0,0) 
		circle [radius=0.16] 
		node[white] {{\fontfamily{qag}\selectfont \tiny \,ID}};
		\draw[white, fill=white] (-0.0525,0.095) 
		circle [radius=0.007];
	\end{tikzpicture}
	\hspace{-3mm}
}
\foreach \x in {A, ..., Z}{\expandafter\xdef\csname orcid\x\endcsname{\noexpand\href{https://orcid.org/\csname orcidauthor\x\endcsname}
		{\noexpand\orcidicon}}
}

\newcommand{\uam}{Department of Theoretical Condensed Matter Physics, Universidad Aut\'onoma de Madrid, 28049 Madrid, Spain}
\newcommand{\ifimac}{Condensed Matter Physics Center (IFIMAC), Universidad Aut\'onoma de Madrid, 28049 Madrid, Spain}
\newcommand{\inc}{Instituto Nicol\'as Cabrera, Universidad Aut\'onoma de Madrid, 28049 Madrid, Spain}
\newcommand{\uppsala}{Department of Physics and Astronomy, Uppsala University, Box 516, S-751 20 Uppsala, Sweden}
\newcommand{\nagoya}{Department of Applied Physics, Nagoya University, Nagoya 464-8603, Japan}

\begin{document}

\title{The Josephson effect in Fibonacci superconductors}

\author{Ignacio Sardinero\orcidC{}}
\affiliation{\uam}
\affiliation{\ifimac}

\author{Jorge Cayao\orcidB{}}
\affiliation{\uppsala}

\author{Keiji Yada}
\affiliation{\nagoya}

\author{Yukio Tanaka\orcidD{}}
\affiliation{\nagoya}

\author{Pablo Burset\orcidA{}}
\affiliation{\uam}
\affiliation{\ifimac}
\affiliation{\inc}

\date{\today}

\begin{abstract}
We theoretically investigate the Josephson effect between two proximized Fibonacci quasicrystals. A quasiperiodic modulation of the chemical potential on a superconducting substrate induces topological gaps and edge modes with energies above the superconducting gap. We reveal that these edge modes develop superconducting correlations which significantly impact the Josephson current, and we term them Fibonacci-Andreev bound states. Notably, the contribution from these edge modes can be controlled by the Fibonacci sequence arrangement, known as phason angle, and can dominate the Josephson effect over the conventional subgap Andreev bound states in short junctions. 
The interplay between the Josephson effect and nontrivial edge modes in quasiperiodic systems presents new opportunities for exploring exotic superconducting phenomena in quasicrystals. 
\end{abstract}
\maketitle

%%%%%%%%%%%%%%%%%%%%%%%%%%%%%%%
% SECTION 1:                 INTRODUCTION                                %
%%%%%%%%%%%%%%%%%%%%%%%%%%%%%%%
\textit{Introduction}.---
The Josephson effect describes the flow of a supercurrent between coupled superconductors~\cite{RevModPhys.51.101,Tinkham} and is at the core of present and future quantum applications that range from superconducting electronics~\cite{Linder_2015,braginski2019superconductor,benito2020hybrid} to quantum computing~\cite{annurevqubit,krantz2019quantum,aguado2020majorana,aguado2020perspective}. Devices exhibiting the Josephson effect are also known as \glspl{jj} and the supercurrent they host originates from the transfer of Cooper pairs due to a finite phase difference between the junction's superconducting order parameters~\cite{RevModPhys.76.411}. It is well established that such supercurrent is carried by \glspl{abs} that emerge within the superconducting gap~\cite{Aslamasov68,kulik69,kulik1975,furusaki1991dc,PhysRevB.45.10563,Furusaki_1999,kashiwaya2000tunnelling,sauls2018andreev}, thus revealing properties of the superconducting state~\cite{tanaka2011symmetry,Kashuba2017May,lutchyn2018majorana,prada2019andreev,frolov2019quest,flensberg2021engineered,Alvarado2023Sep,Sardinero2024,SeoaneSouto2024Oct,fukayaCayaoReview2025,Cayao2020odd}.

When superconductors in \glspl{jj} couple via an intermediate normal region, the behavior of \glspl{abs} is modified according to the properties of such a region~\cite{Beenakker:92,mizushima2018multifaceted,tanaka2024theory}. 
In particular, the length of the bridging region sets the number of \glspl{abs} contributing to the supercurrent and hence determines the Josephson effect~\cite{Beenakker:92}. In the case of \glspl{jj} with a small intermediate region, so called short \glspl{jj}, the supercurrent is determined by only a pair of \glspl{abs}~\cite{RevModPhys.76.411}. This phenomenon has been extensively studied theoretically~\cite{
%PhysRevLett.72.554,PhysRevB.54.7366,
AGRAIT200381,Martin_Rodero_2011,PhysRevLett.108.257001,
PhysRevB.96.024516, cayao2018finite,cayao2018andreev,
PhysRevB.108.075425,PhysRevLett.123.117001,PhysRevB.104.L020501,Zhao2025,PhysRevB.106.L100502,
Yang_2023,Lu2020Jan,Cayao2022Sep,Cayao2024May,PhysRevB.110.L201403,PhysRevB.93.174502,PhysRevB.106.195415,PhysRevB.110.235426,PhysRevB.92.035428} and experimentally~\cite{goffman2000supercurrent,pillet2010andreev,spanton2017current,PhysRevLett.124.226801,PhysRevLett.125.116803,Janvier2015,Tosi2019}, 
becoming the standard paradigm for the Josephson effect. 

\begin{figure}[b]
 \centering
\includegraphics[width=1.0\columnwidth]{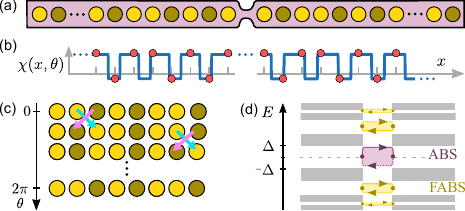}
\caption{
(a) Sketch of a Josephson junction formed by two Fibonacci superconductors with onsite energies following a Fibonacci sequence (yellow/brown circles). 
(b) Spatial profile of the Fibonacci sequence of onsite energies (red dots) generated by $\chi(x,\theta)$ (blue line). 
(c) Example of dislocations caused by the phason angle $\theta$. 
(d) Sketch of the energy spectrum of the Josephson junction showing conventional \gls{abs} (magenta arrows) within the superconducting gap and emergent \glspl{fabs} (yellow arrows) within higher energy Fibonacci gaps. }
 \label{fig1}
\end{figure}

Here we show that this standard picture--where the Josephson effect in short junctions is governed by subgap \glspl{abs}--fails to fully describe the Josephson effect in quasicrystals. 
To demonstrate this, we consider \glspl{jj} formed by Fibonacci superconductors, which correspond to one-dimensional Fibonacci quasicrystals with proximity-induced superconductivity (\cref{fig1}). 
We find that a short \gls{jj} formed by Fibonacci superconductors hosts high energy gaps that bound states, which we term \glspl{fabs}. These \glspl{fabs} inherit the normal-state quasicrystal topology~\cite{Jagannathan_RMP_2021}, which allows them to display a strong dependence on the phase difference between the superconductors forming the junction, so that their contribution can dominate the Josephson effect. In these cases, the conventional \glspl{abs} are suppressed while the \glspl{fabs} localize at the junction; an effect stemming from the topological properties of the parent quasicrystal. 
Our work introduces promising directions for exploring the physics of quasicrystals in superconducting hybrid structures~\cite{Kamiya_NatComm2018,Wang_arxiv_2024,Sandberg_PRB_2024,Kobialka_PRB_2024}.

%%%%%%%%%%%%%%%%%%%%%%%%%%%%%%%
% SECTION 2:      MODEL         %
%%%%%%%%%%%%%%%%%%%%%%%%%%%%%%%
\textit{Fibonacci \glspl{jj}}.---
We consider one-dimensional \glspl{jj} formed by conventional spin-singlet $s$-wave superconductors featuring onsite quasiperiodic energy modulations arranged in a Fibonacci sequence, see \cref{fig1}(a). This Fibonacci \gls{jj} is modeled by 
\begin{equation}
    H = H_L + H_R + H_{T},
    \label{eq:fullJJ}
\end{equation}
where $H_{L(R)}$ describes the left (right) superconductor with $N_{L(R)}$ lattice sites, and is given by ($\alpha=L,R$)
\begin{equation}
 \label{eq:HLR}
	\begin{split}
		H_{\alpha} = &\sum_{i=1}^{ N_{\alpha}} \left( \varepsilon_i  -\mu \right) c_{i\sigma}^\dagger c_{i\sigma} 
		-t\sum_{i=1}^{ N_{\alpha}} c_{i+1\sigma}^\dagger c_{i\sigma}  \\
		 +&   \sum_{i=1}^{ N_{\alpha}} \Delta \me^{i\phi_{\alpha}} c_{i\sigma}^\dagger c_{i\bar{\sigma}}^\dagger  + {\rm H.c.}\,,
	\end{split}
\end{equation}
with fermionic creation (annihilation) operators $c_{i\sigma}$ ($c_{i\sigma}^\dagger$) acting on lattice site $i$ with spin $\sigma=\uparrow,\downarrow$ ($\bar{\sigma}=\downarrow,\uparrow$), while $H_{T} = -t_0 c_{ N_{L}+1}^\dagger c^{}_{ N_{L}} + {\rm H.c.}$ models the coupling between superconductors, with $t_{0}$ being the coupling strength. 
In \cref{eq:HLR}, $t$ is the hopping amplitude, $\mu$ the chemical potential, and $\varepsilon_{i}$ is the space-dependent onsite energy which follows a quasiperiodic profile corresponding to the so-called diagonal Fibonacci chain, which we specify below. 
We choose the same spin-singlet $s$-wave pairing amplitude $\Delta>0$ on both superconductors. The superconducting phases are $\phi_{\alpha}$, and $\phi=\phi_R-\phi_L$ their difference. 

The onsite energies alternate between the values $\varepsilon_{a,b}$, see \cref{fig1}(b), following the quasiperiodic generating function $\chi_j =\mathrm{sgn}[\cos(2\pi j\omega + \theta) - \cos(\pi\omega)]$, with $\varepsilon_j=\varepsilon_{a,b}$ for $\chi_j\gtrless0$~\cite{Jagannathan_RMP_2021}.
Here, $\omega = 2/(1+\sqrt{5})$ is a frequency that is not commensurate with the lattice periodicity~\cite{Kraus_PRL2012b}, determined by lattice constant $a$, and $\theta$ is the phason angle that accounts for a dislocation of two consecutive onsite energies $(\varepsilon_i, \varepsilon_{i+1})$ as sketched in \cref{fig1}(c). 
We define the Fibonacci chain setting the golden ratio $\varepsilon_a/\varepsilon_b = - F_{n-1}/F_{n-2}$, with $F_0=1$, $F_1=2$ and $F_n=F_{n-1}+F_{n-2}$ the Fibonacci numbers. 
Moreover, to maintain the average onsite energy around $\mu$ we choose $\varepsilon_b$ so that $\sum_i \varepsilon_i=0$, leaving only $\varepsilon_a$ as the nontrivial parameter controlling the properties of the quasiperiodic chain. Finally, a chain of length $L_\alpha =N_\alpha a$ is represented by the approximant $S_n$, which accounts for $F_n$ chain sites ($F_{n-1}$ sites with $\varepsilon_a$ and $F_{n-2}$ sites with $\varepsilon_b$) and all configurations given by a phason flip $\theta\in[0,2\pi]$~\cite{SM}. Without loss of generality, we only consider symmetric junctions with $N_L=N_R$. 

One of the main features of normal-state quasiperiodic Fibonacci systems is that they host energy gaps [\cref{fig1}(d)], which we refer to as \textit{Fibonacci gaps}. 
The number of these gaps depends on the approximant $S_n$ of the Fibonacci sequence. However, stable gaps appearing for two consecutive approximants will remain open in longer systems. 
The Fibonacci gaps have a topological origin~\cite{Kraus_PRL2012a,Kraus_PRL2012b} featuring localized edge modes with a dispersion that strongly depends on the phason angle $\theta$. The magnitude of the Fibonacci gaps, however, is not affected by $\theta$. 
Below we explore how Fibonacci gaps, and their topological edge states, contribute to the Josephson effect in a \gls{jj} modeled by \cref{eq:fullJJ} where each Fibonacci superconductor has different phason angles $\theta_{L,R}$. 
 
\begin{figure}[t!]
 \centering
\includegraphics[width=1.0\columnwidth]{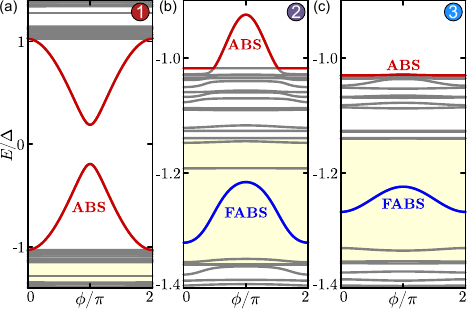}
\caption{Energy spectrum as a function of the superconducting phase difference $\phi$ for fixed $\theta_L$ and $\theta_R = 0.4\pi$ (a), $0.2\pi$ (b), and  $1.8\pi$ (c). 
The red and blue curves correspond to \glspl{abs} and \glspl{fabs}, respectively. A characteristic Fibonacci gap is indicated by yellow shaded areas. Parameters: $\theta_L = 0.9\pi$, $t_0/t = 0.6$, $\varepsilon_a/t = 0.7$, $\mu/t=0$, $\Delta/t=0.1$, and $N_{L,R}=233$ (11-th approximant). 
Cases (a), (b), and (c) are respectively labeled in later figures as \textcircled{\raisebox{-0.5pt}{1}}, \textcircled{\raisebox{-0.5pt}{2}} and \textcircled{\raisebox{-0.5pt}{3}}. 
}
 \label{fig2}
\end{figure}

%%%%%%%%%%%%%%%%%%%%%%%%%%%%%%%
% SECTION 3:      Spectrum         %
%%%%%%%%%%%%%%%%%%%%%%%%%%%%%%%
\textit{Phase-dependent energy spectrum}.---
We begin by obtaining the junction's energy spectrum as a function of the superconducting phase difference for a fixed $\theta_L$ and different values of $\theta_R$, see \cref{fig2}. For $\theta_{L}=0.9\pi$ and $\theta_{R}=0.4\pi$, a pair of \glspl{abs} [red curves in \cref{fig2}(a)] appear within the superconducting gap $\Delta$ featuring a finite energy minigap $\delta_{\pi}\neq0$ at $\phi=\pi$, that reflects a finite, not perfect transmission across the junction. 
A quasicontinuum forms above $\Delta$~\footnote{
Since the spectrum of the Fibonacci chain has zero Lebesgue measure~\cite{Jagannathan_RMP_2021}, there is no continuum of states, even for an infinite chain. We use the term \textit{quasicontinuum} to refer to the fact that edge states in gaps much smaller than $\Delta$ present a negligible contribution to the supercurrent and can thus be ignored. 
}, displaying a series of closely-packed bands interrupted by distinctive Fibonacci gaps. These finite-size gaps originate from the quasiperiodicity of the superconductors and can be either stable or transient~\cite{Jagannathan_RMP_2021}. For example, a stable gap appearing for the $n$-th approximant $S_n$ remains for longer approximants $S_{m>n}$, and is topologically protected~\cite{Kraus_PRL2012b}, while the transient gaps disappear. 
We checked that our results belong to stable gaps~\cite{SM}. The number of levels between two Fibonacci gaps depends on the length of the superconductors. For sufficiently long ones, the quasicontinuum only contains gaps much smaller than $\Delta$ and behaves as a continuum, i.e., it is independent of $\phi$.

As one of the phason angles varies ($\theta_R$) the minigap $\delta_{\pi}$ increases and the \glspl{abs} (red curves) tend to merge with the quasicontinuum and become dispersionless~\cite{SM}. Interestingly, in this regime the quasicontinuum hosts energy levels with a strong dependence on $\phi$, akin to the \glspl{abs} below $\Delta$ but inside the Fibonacci gaps, see shaded yellow regions in \cref{fig2} for a representative example. We thus term these emergent states \glspl{fabs} (blue curves). Under general circumstances both \glspl{abs} and \glspl{fabs} coexist at different energies, as shown in \cref{fig2}(b). However, there are regimes where the \glspl{abs} become dispersionless with respect to $\phi$ while the \glspl{fabs} exhibit the strongest phase dependence, see \cref{fig2}(c). Consequently, a Fibonacci \gls{jj} can host discrete phase-dependent \glspl{fabs} at energies over the superconducting gap $\Delta$, even in the short junction regime. 
This phenomenon expands the conventional framework for short \glspl{jj}, in which the junction hosts only a pair of \glspl{abs} inside the superconducting gap~\cite{SM}. 

\begin{figure}[t!]
 \centering
    \includegraphics[width=1.0\columnwidth]{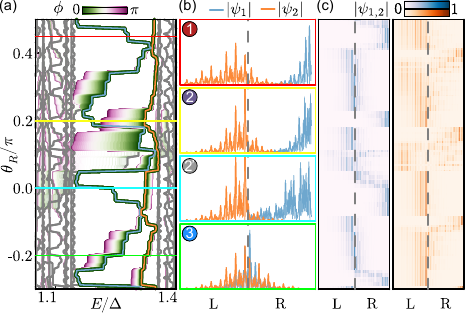}
\caption{
(a) Energy spectrum around a Fibonacci gap as a function of the phason angle $\theta_R$ for different values of the superconducting phase difference $\phi$ indicated by the color bar. 
(b) Modulus of the \glspl{fabs} wavefunction as a function of the position along the junction, for fixed values of the phason angle $\theta_{R}$ in (a) marked by the horizontal colored lines. Cases \textcircled{\raisebox{-0.5pt}{1}}, \textcircled{\raisebox{-0.5pt}{2}}, and \textcircled{\raisebox{-0.5pt}{3}} have same parameters as in \cref{fig2}. 
(c) Modulus of the \glspl{fabs} wavefunctions in (a) as a function of $\theta_{R}$ and space at $\phi=0$. In all cases, $t_0/t = 0.6$, $\varepsilon_a/t = 0.7$, $\mu/t=0$, $\Delta/t=0.1$ and $\theta_L=0.9\pi$. 
}
 \label{fig3}
\end{figure}

We now try to unveil the nontrivial interplay between the superconducting phase difference and the phason angles that gives rise to the \glspl{fabs}. 
We focus on the Fibonacci gap in \cref{fig2}(c) displaying a \gls{fabs} (blue line), and plot in \cref{fig3}(a) the spectrum around it (horizontal axis) as a function of the phason angle $\theta_R$ (vertical axis) and $\phi$ (color lines). 
Since the generating function $\chi_j$ is $2\pi$ periodic in $\theta$, the system is $2\pi$ periodic with each of the phason angles $\theta_{L,R}$. As a result, $\theta_{R}=1.8\pi$ used in \cref{fig2}(c) corresponds to the green line with $\theta_R=-0.2\pi$ in \cref{fig3}(a).  
Inside the Fibonacci gap, the \glspl{fabs} are strongly dependent to variations of both the phason angles ($\theta_{R}$) and the superconducting phase difference. This effect is indeed identified by noting that the \glspl{fabs} change color as $\theta_{R}$ varies, indicating a dependence on $\phi$. 

Notably, depending on the values of phason angles and superconducting phase difference, the \glspl{fabs} can be localized at distinct places in the \gls{jj}. This effect is shown in \cref{fig3}(b), where we plot the square modulus of the wave function of the \glspl{fabs} as a function of position at fixed values of $\theta_{R}$ marked by horizontal lines in \cref{fig3}(a). For instance, by choosing $\theta_{R}=0.42\pi$ in the top panel of \cref{fig3}(b), for $\phi=0$ one of the \glspl{fabs} (blue curve) is located at the outer edge of the right superconductor (R), while the other \glspl{fabs} (orange curve) is extended over the left superconductor (L) with a slightly enhanced weight near the junction. Distinct values of $\theta_{R}$ and $\phi$ strongly affect the localization of the \glspl{fabs}, as observed in the second, third, and four panels of \cref{fig3}(b). Interestingly, when $\theta_{R}=-0.2\pi$ [corresponding to \cref{fig2}(c)] and $\phi=0$, the \gls{fabs} located at the edge of R in the top panel (blue curve) becomes located at the junction, a localization effect that also happens for the second \gls{fabs} (orange curve). The localization effects are further explored in \cref{fig3}(c), where we plot the square modulus of the wave function of the two \glspl{fabs} $|\psi_{1,2}|^{2}$ in (b) as a function of $\theta_{R}$ and space at $\phi=0$. Here, the highest intensity of $|\psi_{1,2}|^{2}$ reveals the localization of both \glspl{fabs}. The localization of \glspl{fabs} in the \gls{jj}, which entirely stems from the quasicrystal topology, is thus highly controllable by the interplay of phason angles and superconducting phase difference. 

\begin{figure}[t!]
 \centering
    \includegraphics[width=1.0\columnwidth]{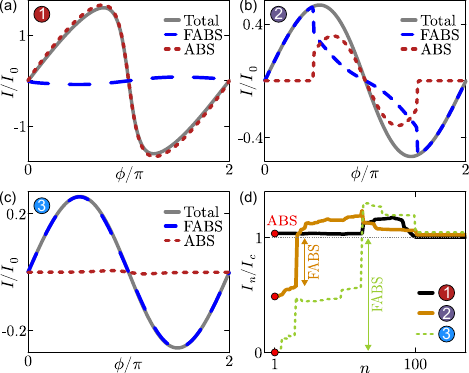}
\caption{
(a-c) Supercurrent as a function of the superconducting phase difference $\phi$ at phason angles that correspond to the energy spectra of \cref{fig2}, $I_0=e\Delta/\hbar$.
(d) For the same cases, maximum value of the supercurrent summed up to the $n$-th eigenstate, $I_n$, as a function of $n$. Red dots mark the contribution from \gls{abs}, $I_1$. Same parameters as in \cref{fig2}.}
 \label{fig4}
\end{figure}

%%%%%%%%%%%%%%%%%%%%%%%%%%%%%%%
% SECTION 4:   SUPERCURRENTS   %
%%%%%%%%%%%%%%%%%%%%%%%%%%%%%%%
\textit{Current-phase characteristics}.---
Having demonstrated the emergence of \glspl{fabs}, we now investigate their impact on the Josephson effect. For this purpose we use the phase-dependent energy spectrum $E_{n}$ discussed before and calculate the zero-temperature supercurrent as~\cite{PhysRevB.96.205425,cayao2018andreev,Sardinero2024} 
\begin{equation}\label{eq:current}
    I(\phi)= - \frac{2e}{\hbar} \left( \frac{ \md E_{1} }{ \md \phi } 
    + \sum_{n>1} \frac{ \md E_{n} }{ \md \phi } \right) 
    \equiv I_\text{ABS} + I_\text{FABS} ,
\end{equation}
where we are identifying the contribution from subgap \glspl{abs} from that of the \glspl{fabs} above $\Delta$. 
In \cref{fig4}(a-c) we present the supercurrents (gray lines) as a function of the superconducting phase difference $I(\phi)$ for the three energy spectra cases presented in \cref{fig2}(a-c). We indicate the contributions from \glspl{abs}, $I_\text{ABS}$ dotted red lines, and \glspl{fabs}, $I_\text{FABS}$ blue dashed lines. 

When \glspl{abs} dominate and exhibit the strongest phase dependence [\cref{fig2}(a)], the total supercurrent is almost entirely given by \gls{abs} contributions [\cref{fig4}(a)].  
For phason angles such that \glspl{fabs} and \glspl{abs} exhibit a similar dependence on the superconducting phase difference, but at different energies, the total supercurrent is determined by the combined interplay of both \glspl{abs} and \glspl{fabs}, see \cref{fig4}(b). 
Indeed, as the \glspl{abs} merge with the quasicontinuum, only their contribution around $\phi\sim\pi$ remains. By contrast, the \glspl{fabs} inside well-defined Fibonacci gaps feature a finite contribution for all phases, cf. \cref{fig2}(b). 
For phason angles where the \glspl{abs} are dispersionless with $\phi$ [\cref{fig2}(c)], the total supercurrent is dominated by \glspl{fabs}, as depicted in \cref{fig4}(c). 
The current-phase relation in the \gls{abs} dominated case acquires a non-sinusoidal behavior [\cref{fig4}(a)], characteristic for highly-transmitting \glspl{jj} with conventional superconductors~\cite{Beenakker:92,tanaka2024theory}.
By contrast, as the supercurrent becomes dominated by contributions from \glspl{fabs} the current-phase relation is more sinusoidal, as corresponds to tunnel junctions. 

A supercurrent dominated by \glspl{fabs} is thus a direct result of a quasiperiodic potential breaking the energy continuum into small gaps, with some hosting bound states at energies above $\Delta$. 
The junction transmission, and thus the dispersion of the conventional \glspl{abs} it hosts, is also affected by the quasiperiodic potential~\cite{SM}. 
As \cref{fig4}(c) demonstrates, the phason angles associated to the quasicrystal topological nature of the Fibonacci \glspl{jj} can render the \glspl{abs} dispersionless with $\phi$ and with vanishing supercurrent contribution. 
Interestingly, in this regime the \glspl{fabs} emerging inside higher energy Fibonacci gaps acquire the strongest dependence on $\phi$ and determine the entire Josephson effect. As a result, the Josephson effect originates from the interplay between the superconducting phase difference and the phason angles, which determine the arrangement of the Fibonacci quasicrystal and unveil its normal-state topology~\cite{SM}. 

We can further investigate the supercurrent contribution due to \glspl{fabs}, and that these states belong within the Fibonacci gaps. We define 
\begin{equation}\label{eq:In}
    I_n = \left( \sum_{i=1}^n \frac{\md E_{i}} {\md\phi}  \right)_{\phi=\phi_c}
\end{equation}
with $\phi_c$ the phase of the critical current. $I_n$ thus represents the contribution to the critical current $I_c$ from the first $n$ states below the Fermi energy. 
As we consider all contributions, $I_n$ converges to the critical current, $I_{n\gg1}\to I_c$. 

In \cref{fig4}(d), we plot $I_{n}$ normalized to the critical current $I_c$ for each of the three current-phase curves presented in \cref{fig4}(a-c). For the current in \cref{fig4}(a) where \glspl{abs} dominate (black line), the first state $I_1$ corresponding to the \gls{abs} (red dot) with energy within the superconducting gap $\Delta$ almost completely dominates the current. We also have that $I_1\gtrsim I_c$, but only fully converges to the critical current after taking into account a large number of states from the quasicontinuum. 

The current in \cref{fig4}(b) features contributions form both \glspl{fabs} and \glspl{abs}. Consequently, $I_1$ is already an important fraction of the critical current (red dot on brown line), while the next 10-15 states provide the remaining important contributions. It is now clearer than in the previous case that not all the contributions are positive, since, e.g., $I_{50}>I_c$, but $I_{100}\simeq I_c$~\footnote{
We note that for finite-size \glspl{jj} there is a finite contribution to the current from states above the gap that is only canceled out after considering all the energy levels. This effect explains the values $I_n/I_c>1$ for finite $n$, which do not mean that the current increases over the maximum current that flows across the junction. 
}. 
We remark that almost all contributions for $n>1$ correspond to states above the superconducting gap $\Delta$. For the third case with dominant \glspl{fabs}, \cref{fig4}(c), the contribution from the \gls{abs} is almost negligible (red dot on the green line), i.e., $I_1\sim0$. Now, not one but several \glspl{fabs} contribute to the total current, some of them at high $n$ values corresponding to Fibonacci gaps at energies much greater than $\Delta$. 
Since at such higher energies the only levels dispersing with the superconducting phase difference are the \glspl{fabs}, the current depicted in green in \cref{fig4}(d) supports that the Josephson effect is entirely driven by the \glspl{fabs}~\cite{SM}.

%%%%%%%%%%%%%%%%%%%%%%%%%%%%%%%
%     CONCLUSIONS           %
%%%%%%%%%%%%%%%%%%%%%%%%%%%%%%%
\textit{Conclusions}.---
We have demonstrated the presence of additional Andreev bound states in the spectrum of Fibonacci Josephson junctions. 
These states appear for energies above the superconducting gap due to the breakdown of the continuum by the quasiperiodic potential, and their localization at the edges of the Fibonacci superconductors is determined by the phason parameter. 
This phason degree of freedom controls the modulation of the quasiperiodic potential, directly affecting the effective transparency of the junction and the relative amplitude of the conventional subgap and Fibonacci over-the-gap Andreev states. 
The emergent Fibonacci Andreev bound states develop a strong phase-dependence when localized at the junction interface, which can make them more prominent than the subgap Andreev bound states. 
In this situation, the Fibonacci states have the dominant contribution to the supercurrent, surpassing the contribution of the subgap Andreev bound state. 
These findings open new avenues for exploring the physics of quasicrystalline superconducting systems, whether they appear naturally~\cite{Kamiya_NatComm2018} or are engineered~\cite{Dvir2023Feb,Bordin_PRL2024,Zatelli2024Sep}. In particular, recent experimental advances on tailored systems of quantum dots~\cite{Bordin_PRL2024} or local voltage gates~\cite{Tombros_JApplPhys2011May,Buscema_Nature2014,Maurand_JCarbon2014} could be a pathway for controlling phason angles in experiments. 

%%%%%%%%%%%%%%%%%%%%%%%%%%%%%%%
%    ACKNOWLEDGMENTS          %
%%%%%%%%%%%%%%%%%%%%%%%%%%%%%%%

I.S. and P.B. acknowledge support from the Spanish CM ``Talento Program'' project No.~2019-T1/IND-14088 and No.~2023-5A/IND-28927, the Agencia Estatal de Investigaci\'on project No.~PID2020-117992GA-I00 and No.~CNS2022-135950 and through the ``María de Maeztu'' Programme for Units of Excellence in R\&D (CEX2023-001316-M). 
J.C. acknowledges the financial support from the Swedish Research Council (Vetenskapsr{\aa}det Grant No. 2021-04121) and the Carl Trygger's Foundation (Grant No. 22: 2093). 
Y.T. acknowledges financial support from JSPS with Grants-in-Aid for Scientific Research (KAKENHI Grants Nos. 23K17668, 24K00583, 24K00556, and 24K00578).

%\bibliography{bibliography}

\begin{thebibliography}{79}%
\makeatletter
\providecommand \@ifxundefined [1]{%
 \@ifx{#1\undefined}
}%
\providecommand \@ifnum [1]{%
 \ifnum #1\expandafter \@firstoftwo
 \else \expandafter \@secondoftwo
 \fi
}%
\providecommand \@ifx [1]{%
 \ifx #1\expandafter \@firstoftwo
 \else \expandafter \@secondoftwo
 \fi
}%
\providecommand \natexlab [1]{#1}%
\providecommand \enquote  [1]{``#1''}%
\providecommand \bibnamefont  [1]{#1}%
\providecommand \bibfnamefont [1]{#1}%
\providecommand \citenamefont [1]{#1}%
\providecommand \href@noop [0]{\@secondoftwo}%
\providecommand \href [0]{\begingroup \@sanitize@url \@href}%
\providecommand \@href[1]{\@@startlink{#1}\@@href}%
\providecommand \@@href[1]{\endgroup#1\@@endlink}%
\providecommand \@sanitize@url [0]{\catcode `\\12\catcode `\$12\catcode
  `\&12\catcode `\#12\catcode `\^12\catcode `\_12\catcode `\%12\relax}%
\providecommand \@@startlink[1]{}%
\providecommand \@@endlink[0]{}%
\providecommand \url  [0]{\begingroup\@sanitize@url \@url }%
\providecommand \@url [1]{\endgroup\@href {#1}{\urlprefix }}%
\providecommand \urlprefix  [0]{URL }%
\providecommand \Eprint [0]{\href }%
\providecommand \doibase [0]{https://doi.org/}%
\providecommand \selectlanguage [0]{\@gobble}%
\providecommand \bibinfo  [0]{\@secondoftwo}%
\providecommand \bibfield  [0]{\@secondoftwo}%
\providecommand \translation [1]{[#1]}%
\providecommand \BibitemOpen [0]{}%
\providecommand \bibitemStop [0]{}%
\providecommand \bibitemNoStop [0]{.\EOS\space}%
\providecommand \EOS [0]{\spacefactor3000\relax}%
\providecommand \BibitemShut  [1]{\csname bibitem#1\endcsname}%
\let\auto@bib@innerbib\@empty
%</preamble>
\bibitem [{\citenamefont {Likharev}(1979)}]{RevModPhys.51.101}%
  \BibitemOpen
  \bibfield  {author} {\bibinfo {author} {\bibfnamefont {K.~K.}\ \bibnamefont
  {Likharev}},\ }\bibfield  {title} {\bibinfo {title} {Superconducting weak
  links},\ }\href {https://doi.org/10.1103/RevModPhys.51.101} {\bibfield
  {journal} {\bibinfo  {journal} {Rev. Mod. Phys.}\ }\textbf {\bibinfo {volume}
  {51}},\ \bibinfo {pages} {101} (\bibinfo {year} {1979})}\BibitemShut
  {NoStop}%
\bibitem [{\citenamefont {Tinkham}(2004)}]{Tinkham}%
  \BibitemOpen
  \bibfield  {author} {\bibinfo {author} {\bibfnamefont {M.}~\bibnamefont
  {Tinkham}},\ }\href@noop {} {\emph {\bibinfo {title} {Introduction to
  superconductivity}}}\ (\bibinfo  {publisher} {Courier Corporation},\ \bibinfo
  {year} {2004})\BibitemShut {NoStop}%
\bibitem [{\citenamefont {Linder}\ and\ \citenamefont
  {Robinson}(2015)}]{Linder_2015}%
  \BibitemOpen
  \bibfield  {author} {\bibinfo {author} {\bibfnamefont {J.}~\bibnamefont
  {Linder}}\ and\ \bibinfo {author} {\bibfnamefont {J.~W.~A.}\ \bibnamefont
  {Robinson}},\ }\bibfield  {title} {\bibinfo {title} {Superconducting
  spintronics},\ }\href {https://doi.org/10.1038/nphys3242} {\bibfield
  {journal} {\bibinfo  {journal} {Nature Physics}\ }\textbf {\bibinfo {volume}
  {11}},\ \bibinfo {pages} {307–315} (\bibinfo {year} {2015})}\BibitemShut
  {NoStop}%
\bibitem [{\citenamefont {Braginski}(2019)}]{braginski2019superconductor}%
  \BibitemOpen
  \bibfield  {author} {\bibinfo {author} {\bibfnamefont {A.~I.}\ \bibnamefont
  {Braginski}},\ }\bibfield  {title} {\bibinfo {title} {{Superconductor
  electronics: Status and outlook}},\ }\href
  {https://doi.org/10.1007/s10948-018-4884-4} {\bibfield  {journal} {\bibinfo
  {journal} {J. Supercond. Novel Magn.}\ }\textbf {\bibinfo {volume} {32}},\
  \bibinfo {pages} {23} (\bibinfo {year} {2019})}\BibitemShut {NoStop}%
\bibitem [{\citenamefont {Benito}\ and\ \citenamefont
  {Burkard}(2020)}]{benito2020hybrid}%
  \BibitemOpen
  \bibfield  {author} {\bibinfo {author} {\bibfnamefont {M.}~\bibnamefont
  {Benito}}\ and\ \bibinfo {author} {\bibfnamefont {G.}~\bibnamefont
  {Burkard}},\ }\bibfield  {title} {\bibinfo {title} {Hybrid
  superconductor-semiconductor systems for quantum technology},\ }\href
  {https://doi.org/10.1063/5.0004777} {\bibfield  {journal} {\bibinfo
  {journal} {Applied Physics Letters}\ }\textbf {\bibinfo {volume} {116}},\
  \bibinfo {pages} {190502} (\bibinfo {year} {2020})}\BibitemShut {NoStop}%
\bibitem [{\citenamefont {Kjaergaard}\ \emph {et~al.}(2020)\citenamefont
  {Kjaergaard}, \citenamefont {Schwartz}, \citenamefont {Braumüller},
  \citenamefont {Krantz}, \citenamefont {Wang}, \citenamefont {Gustavsson},\
  and\ \citenamefont {Oliver}}]{annurevqubit}%
  \BibitemOpen
  \bibfield  {author} {\bibinfo {author} {\bibfnamefont {M.}~\bibnamefont
  {Kjaergaard}}, \bibinfo {author} {\bibfnamefont {M.~E.}\ \bibnamefont
  {Schwartz}}, \bibinfo {author} {\bibfnamefont {J.}~\bibnamefont
  {Braumüller}}, \bibinfo {author} {\bibfnamefont {P.}~\bibnamefont {Krantz}},
  \bibinfo {author} {\bibfnamefont {J.~I.-J.}\ \bibnamefont {Wang}}, \bibinfo
  {author} {\bibfnamefont {S.}~\bibnamefont {Gustavsson}},\ and\ \bibinfo
  {author} {\bibfnamefont {W.~D.}\ \bibnamefont {Oliver}},\ }\bibfield  {title}
  {\bibinfo {title} {{Superconducting Qubits: Current State of Play}},\ }\href
  {https://doi.org/10.1146/annurev-conmatphys-031119-050605} {\bibfield
  {journal} {\bibinfo  {journal} {Annu. Rev. Condens. Matter Phys.}\ ,\
  \bibinfo {pages} {369}} (\bibinfo {year} {2020})}\BibitemShut {NoStop}%
\bibitem [{\citenamefont {Krantz}\ \emph {et~al.}(2019)\citenamefont {Krantz},
  \citenamefont {Kjaergaard}, \citenamefont {Yan}, \citenamefont {Orlando},
  \citenamefont {Gustavsson},\ and\ \citenamefont
  {Oliver}}]{krantz2019quantum}%
  \BibitemOpen
  \bibfield  {author} {\bibinfo {author} {\bibfnamefont {P.}~\bibnamefont
  {Krantz}}, \bibinfo {author} {\bibfnamefont {M.}~\bibnamefont {Kjaergaard}},
  \bibinfo {author} {\bibfnamefont {F.}~\bibnamefont {Yan}}, \bibinfo {author}
  {\bibfnamefont {T.~P.}\ \bibnamefont {Orlando}}, \bibinfo {author}
  {\bibfnamefont {S.}~\bibnamefont {Gustavsson}},\ and\ \bibinfo {author}
  {\bibfnamefont {W.~D.}\ \bibnamefont {Oliver}},\ }\bibfield  {title}
  {\bibinfo {title} {A quantum engineer's guide to superconducting qubits},\
  }\href {https://doi.org/10.1063/1.5089550} {\bibfield  {journal} {\bibinfo
  {journal} {Appl. Phys. Rev.}\ }\textbf {\bibinfo {volume} {6}},\ \bibinfo
  {pages} {021318} (\bibinfo {year} {2019})}\BibitemShut {NoStop}%
\bibitem [{\citenamefont {Aguado}\ and\ \citenamefont
  {Kouwenhoven}(2020)}]{aguado2020majorana}%
  \BibitemOpen
  \bibfield  {author} {\bibinfo {author} {\bibfnamefont {R.}~\bibnamefont
  {Aguado}}\ and\ \bibinfo {author} {\bibfnamefont {L.~P.}\ \bibnamefont
  {Kouwenhoven}},\ }\bibfield  {title} {\bibinfo {title} {{Majorana qubits for
  topological quantum computing}},\ }\href {https://doi.org/10.1063/PT.3.4499}
  {\bibfield  {journal} {\bibinfo  {journal} {Phys. Today}\ }\textbf {\bibinfo
  {volume} {73}},\ \bibinfo {pages} {44} (\bibinfo {year} {2020})}\BibitemShut
  {NoStop}%
\bibitem [{\citenamefont {Aguado}(2020)}]{aguado2020perspective}%
  \BibitemOpen
  \bibfield  {author} {\bibinfo {author} {\bibfnamefont {R.}~\bibnamefont
  {Aguado}},\ }\bibfield  {title} {\bibinfo {title} {A perspective on
  semiconductor-based superconducting qubits},\ }\href
  {https://doi.org/10.1063/5.0024124} {\bibfield  {journal} {\bibinfo
  {journal} {Applied Physics Letters}\ }\textbf {\bibinfo {volume} {117}},\
  \bibinfo {pages} {240501} (\bibinfo {year} {2020})}\BibitemShut {NoStop}%
\bibitem [{\citenamefont {Golubov}\ \emph {et~al.}(2004)\citenamefont
  {Golubov}, \citenamefont {Kupriyanov},\ and\ \citenamefont
  {Il'ichev}}]{RevModPhys.76.411}%
  \BibitemOpen
  \bibfield  {author} {\bibinfo {author} {\bibfnamefont {A.~A.}\ \bibnamefont
  {Golubov}}, \bibinfo {author} {\bibfnamefont {M.~Y.}\ \bibnamefont
  {Kupriyanov}},\ and\ \bibinfo {author} {\bibfnamefont {E.}~\bibnamefont
  {Il'ichev}},\ }\bibfield  {title} {\bibinfo {title} {The current-phase
  relation in {J}osephson junctions},\ }\href
  {https://doi.org/10.1103/RevModPhys.76.411} {\bibfield  {journal} {\bibinfo
  {journal} {Rev. Mod. Phys.}\ }\textbf {\bibinfo {volume} {76}},\ \bibinfo
  {pages} {411} (\bibinfo {year} {2004})}\BibitemShut {NoStop}%
\bibitem [{\citenamefont {Aslamasov}\ \emph {et~al.}(1968)\citenamefont
  {Aslamasov}, \citenamefont {Larkin},\ and\ \citenamefont
  {Ovchinnikov}}]{Aslamasov68}%
  \BibitemOpen
  \bibfield  {author} {\bibinfo {author} {\bibfnamefont {L.~G.}\ \bibnamefont
  {Aslamasov}}, \bibinfo {author} {\bibfnamefont {A.~I.}\ \bibnamefont
  {Larkin}},\ and\ \bibinfo {author} {\bibfnamefont {Y.~N.}\ \bibnamefont
  {Ovchinnikov}},\ }\href@noop {} {\bibfield  {journal} {\bibinfo  {journal}
  {Zh. Eksp. Teor. Fiz.}\ }\textbf {\bibinfo {volume} {55}},\ \bibinfo {pages}
  {323} (\bibinfo {year} {1968})},\ \bibinfo {note} {sov. Phys. JETP 28, 171
  (1969)}\BibitemShut {NoStop}%
\bibitem [{\citenamefont {Kulik}(1969)}]{kulik69}%
  \BibitemOpen
  \bibfield  {author} {\bibinfo {author} {\bibfnamefont {I.~O.}\ \bibnamefont
  {Kulik}},\ }\href@noop {} {\bibfield  {journal} {\bibinfo  {journal} {Zh.
  Eksp. Teor. Fiz.}\ }\textbf {\bibinfo {volume} {57}},\ \bibinfo {pages}
  {1745} (\bibinfo {year} {1969})},\ \bibinfo {note} {sov. Phys. JETP 30, 944
  (1970)}\BibitemShut {NoStop}%
\bibitem [{\citenamefont {Kulik}\ and\ \citenamefont
  {Omel'Yanchuk}(1975)}]{kulik1975}%
  \BibitemOpen
  \bibfield  {author} {\bibinfo {author} {\bibfnamefont {I.}~\bibnamefont
  {Kulik}}\ and\ \bibinfo {author} {\bibfnamefont {A.}~\bibnamefont
  {Omel'Yanchuk}},\ }\bibfield  {title} {\bibinfo {title} {Contribution to the
  microscopic theory of the {J}osephson effect in superconducting bridges},\
  }\href@noop {} {\bibfield  {journal} {\bibinfo  {journal} {JETP Lett.}\
  }\textbf {\bibinfo {volume} {21}},\ \bibinfo {pages} {216} (\bibinfo {year}
  {1975})}\BibitemShut {NoStop}%
\bibitem [{\citenamefont {Furusaki}\ and\ \citenamefont
  {Tsukada}(1991)}]{furusaki1991dc}%
  \BibitemOpen
  \bibfield  {author} {\bibinfo {author} {\bibfnamefont {A.}~\bibnamefont
  {Furusaki}}\ and\ \bibinfo {author} {\bibfnamefont {M.}~\bibnamefont
  {Tsukada}},\ }\bibfield  {title} {\bibinfo {title} {{Dc Josephson effect and
  Andreev reflection}},\ }\href {https://doi.org/10.1016/0038-1098(91)90201-6}
  {\bibfield  {journal} {\bibinfo  {journal} {Solid State Commun.}\ }\textbf
  {\bibinfo {volume} {78}},\ \bibinfo {pages} {299} (\bibinfo {year}
  {1991})}\BibitemShut {NoStop}%
\bibitem [{\citenamefont {Furusaki}\ \emph {et~al.}(1992)\citenamefont
  {Furusaki}, \citenamefont {Takayanagi},\ and\ \citenamefont
  {Tsukada}}]{PhysRevB.45.10563}%
  \BibitemOpen
  \bibfield  {author} {\bibinfo {author} {\bibfnamefont {A.}~\bibnamefont
  {Furusaki}}, \bibinfo {author} {\bibfnamefont {H.}~\bibnamefont
  {Takayanagi}},\ and\ \bibinfo {author} {\bibfnamefont {M.}~\bibnamefont
  {Tsukada}},\ }\bibfield  {title} {\bibinfo {title} {Josephson effect of the
  superconducting quantum point contact},\ }\href
  {https://doi.org/10.1103/PhysRevB.45.10563} {\bibfield  {journal} {\bibinfo
  {journal} {Phys. Rev. B}\ }\textbf {\bibinfo {volume} {45}},\ \bibinfo
  {pages} {10563} (\bibinfo {year} {1992})}\BibitemShut {NoStop}%
\bibitem [{\citenamefont {Furusaki}(1999)}]{Furusaki_1999}%
  \BibitemOpen
  \bibfield  {author} {\bibinfo {author} {\bibfnamefont {A.}~\bibnamefont
  {Furusaki}},\ }\bibfield  {title} {\bibinfo {title} {{Josephson current
  carried by Andreev levels in superconducting quantum point contacts}},\
  }\href {https://doi.org/10.1006/spmi.1999.0730} {\bibfield  {journal}
  {\bibinfo  {journal} {Superlattices Microstruct.}\ }\textbf {\bibinfo
  {volume} {25}},\ \bibinfo {pages} {809} (\bibinfo {year} {1999})}\BibitemShut
  {NoStop}%
\bibitem [{\citenamefont {Kashiwaya}\ and\ \citenamefont
  {Tanaka}(2000)}]{kashiwaya2000tunnelling}%
  \BibitemOpen
  \bibfield  {author} {\bibinfo {author} {\bibfnamefont {S.}~\bibnamefont
  {Kashiwaya}}\ and\ \bibinfo {author} {\bibfnamefont {Y.}~\bibnamefont
  {Tanaka}},\ }\bibfield  {title} {\bibinfo {title} {{Tunnelling effects on
  surface bound states in unconventional superconductors}},\ }\href
  {https://doi.org/10.1088/0034-4885/63/10/202} {\bibfield  {journal} {\bibinfo
   {journal} {Rep. Prog. Phys.}\ }\textbf {\bibinfo {volume} {63}},\ \bibinfo
  {pages} {1641} (\bibinfo {year} {2000})}\BibitemShut {NoStop}%
\bibitem [{\citenamefont {Sauls}(2018)}]{sauls2018andreev}%
  \BibitemOpen
  \bibfield  {author} {\bibinfo {author} {\bibfnamefont {J.~A.}\ \bibnamefont
  {Sauls}},\ }\bibfield  {title} {\bibinfo {title} {Andreev bound states and
  their signatures},\ }\href {https://doi.org/10.1098/rsta.2018.0140}
  {\bibfield  {journal} {\bibinfo  {journal} {Philos. Trans. Royal Soc. A}\
  }\textbf {\bibinfo {volume} {376}},\ \bibinfo {pages} {20180140} (\bibinfo
  {year} {2018})}\BibitemShut {NoStop}%
\bibitem [{\citenamefont {Tanaka}\ \emph {et~al.}(2012)\citenamefont {Tanaka},
  \citenamefont {Sato},\ and\ \citenamefont {Nagaosa}}]{tanaka2011symmetry}%
  \BibitemOpen
  \bibfield  {author} {\bibinfo {author} {\bibfnamefont {Y.}~\bibnamefont
  {Tanaka}}, \bibinfo {author} {\bibfnamefont {M.}~\bibnamefont {Sato}},\ and\
  \bibinfo {author} {\bibfnamefont {N.}~\bibnamefont {Nagaosa}},\ }\bibfield
  {title} {\bibinfo {title} {Symmetry and topology in
  superconductors--odd-frequency pairing and edge states--},\ }\href
  {https://doi.org/10.1143/JPSJ.81.011013} {\bibfield  {journal} {\bibinfo
  {journal} {J. Phys. Soc. Japan}\ }\textbf {\bibinfo {volume} {81}},\ \bibinfo
  {pages} {011013} (\bibinfo {year} {2012})}\BibitemShut {NoStop}%
\bibitem [{\citenamefont {Kashuba}\ \emph {et~al.}(2017)\citenamefont
  {Kashuba}, \citenamefont {Sothmann}, \citenamefont {Burset},\ and\
  \citenamefont {Trauzettel}}]{Kashuba2017May}%
  \BibitemOpen
  \bibfield  {author} {\bibinfo {author} {\bibfnamefont {O.}~\bibnamefont
  {Kashuba}}, \bibinfo {author} {\bibfnamefont {B.}~\bibnamefont {Sothmann}},
  \bibinfo {author} {\bibfnamefont {P.}~\bibnamefont {Burset}},\ and\ \bibinfo
  {author} {\bibfnamefont {B.}~\bibnamefont {Trauzettel}},\ }\bibfield  {title}
  {\bibinfo {title} {Majorana {STM} as a perfect detector of odd-frequency
  superconductivity},\ }\href {https://doi.org/10.1103/PhysRevB.95.174516}
  {\bibfield  {journal} {\bibinfo  {journal} {Phys. Rev. B}\ }\textbf {\bibinfo
  {volume} {95}},\ \bibinfo {pages} {174516} (\bibinfo {year}
  {2017})}\BibitemShut {NoStop}%
\bibitem [{\citenamefont {Lutchyn}\ \emph {et~al.}(2018)\citenamefont
  {Lutchyn}, \citenamefont {Bakkers}, \citenamefont {Kouwenhoven},
  \citenamefont {Krogstrup}, \citenamefont {Marcus},\ and\ \citenamefont
  {Oreg}}]{lutchyn2018majorana}%
  \BibitemOpen
  \bibfield  {author} {\bibinfo {author} {\bibfnamefont {R.~M.}\ \bibnamefont
  {Lutchyn}}, \bibinfo {author} {\bibfnamefont {E.~P.}\ \bibnamefont
  {Bakkers}}, \bibinfo {author} {\bibfnamefont {L.~P.}\ \bibnamefont
  {Kouwenhoven}}, \bibinfo {author} {\bibfnamefont {P.}~\bibnamefont
  {Krogstrup}}, \bibinfo {author} {\bibfnamefont {C.~M.}\ \bibnamefont
  {Marcus}},\ and\ \bibinfo {author} {\bibfnamefont {Y.}~\bibnamefont {Oreg}},\
  }\bibfield  {title} {\bibinfo {title} {Majorana zero modes in
  superconductor--semiconductor heterostructures},\ }\href
  {https://doi.org/10.1038/s41578-018-0003-1} {\bibfield  {journal} {\bibinfo
  {journal} {Nat. Rev. Mater.}\ }\textbf {\bibinfo {volume} {3}},\ \bibinfo
  {pages} {52} (\bibinfo {year} {2018})}\BibitemShut {NoStop}%
\bibitem [{\citenamefont {Prada}\ \emph {et~al.}(2020)\citenamefont {Prada},
  \citenamefont {San-Jose}, \citenamefont {de~Moor}, \citenamefont {Geresdi},
  \citenamefont {Lee}, \citenamefont {Klinovaja}, \citenamefont {Loss},
  \citenamefont {Nyg{\aa}rd}, \citenamefont {Aguado},\ and\ \citenamefont
  {Kouwenhoven}}]{prada2019andreev}%
  \BibitemOpen
  \bibfield  {author} {\bibinfo {author} {\bibfnamefont {E.}~\bibnamefont
  {Prada}}, \bibinfo {author} {\bibfnamefont {P.}~\bibnamefont {San-Jose}},
  \bibinfo {author} {\bibfnamefont {M.~W.~A.}\ \bibnamefont {de~Moor}},
  \bibinfo {author} {\bibfnamefont {A.}~\bibnamefont {Geresdi}}, \bibinfo
  {author} {\bibfnamefont {E.~J.~H.}\ \bibnamefont {Lee}}, \bibinfo {author}
  {\bibfnamefont {J.}~\bibnamefont {Klinovaja}}, \bibinfo {author}
  {\bibfnamefont {D.}~\bibnamefont {Loss}}, \bibinfo {author} {\bibfnamefont
  {J.}~\bibnamefont {Nyg{\aa}rd}}, \bibinfo {author} {\bibfnamefont
  {R.}~\bibnamefont {Aguado}},\ and\ \bibinfo {author} {\bibfnamefont {L.~P.}\
  \bibnamefont {Kouwenhoven}},\ }\bibfield  {title} {\bibinfo {title} {{From
  Andreev to Majorana bound states in hybrid
  superconductor{\textendash}semiconductor nanowires}},\ }\href
  {https://doi.org/10.1038/s42254-020-0228-y} {\bibfield  {journal} {\bibinfo
  {journal} {Nat. Rev. Phys.}\ }\textbf {\bibinfo {volume} {2}},\ \bibinfo
  {pages} {575} (\bibinfo {year} {2020})}\BibitemShut {NoStop}%
\bibitem [{\citenamefont {Frolov}\ \emph {et~al.}(2020)\citenamefont {Frolov},
  \citenamefont {Manfra},\ and\ \citenamefont {Sau}}]{frolov2019quest}%
  \BibitemOpen
  \bibfield  {author} {\bibinfo {author} {\bibfnamefont {S.~M.}\ \bibnamefont
  {Frolov}}, \bibinfo {author} {\bibfnamefont {M.~J.}\ \bibnamefont {Manfra}},\
  and\ \bibinfo {author} {\bibfnamefont {J.~D.}\ \bibnamefont {Sau}},\
  }\bibfield  {title} {\bibinfo {title} {Topological superconductivity in
  hybrid devices},\ }\href {https://doi.org/10.1038/s41567-020-0925-6}
  {\bibfield  {journal} {\bibinfo  {journal} {Nat. Phys.}\ }\textbf {\bibinfo
  {volume} {16}},\ \bibinfo {pages} {718} (\bibinfo {year} {2020})}\BibitemShut
  {NoStop}%
\bibitem [{\citenamefont {Flensberg}\ \emph {et~al.}(2021)\citenamefont
  {Flensberg}, \citenamefont {von Oppen},\ and\ \citenamefont
  {Stern}}]{flensberg2021engineered}%
  \BibitemOpen
  \bibfield  {author} {\bibinfo {author} {\bibfnamefont {K.}~\bibnamefont
  {Flensberg}}, \bibinfo {author} {\bibfnamefont {F.}~\bibnamefont {von
  Oppen}},\ and\ \bibinfo {author} {\bibfnamefont {A.}~\bibnamefont {Stern}},\
  }\bibfield  {title} {\bibinfo {title} {Engineered platforms for topological
  superconductivity and {Majorana} zero modes},\ }\href
  {https://www.nature.com/articles/s41578-021-00336-6} {\bibfield  {journal}
  {\bibinfo  {journal} {Nat. Rev. Mater.}\ }\textbf {\bibinfo {volume} {6}},\
  \bibinfo {pages} {944} (\bibinfo {year} {2021})}\BibitemShut {NoStop}%
\bibitem [{\citenamefont {Alvarado}\ \emph {et~al.}(2023)\citenamefont
  {Alvarado}, \citenamefont {Burset},\ and\ \citenamefont
  {Yeyati}}]{Alvarado2023Sep}%
  \BibitemOpen
  \bibfield  {author} {\bibinfo {author} {\bibfnamefont {M.}~\bibnamefont
  {Alvarado}}, \bibinfo {author} {\bibfnamefont {P.}~\bibnamefont {Burset}},\
  and\ \bibinfo {author} {\bibfnamefont {A.~L.}\ \bibnamefont {Yeyati}},\
  }\bibfield  {title} {\bibinfo {title} {Intrinsic nonmagnetic
  ${\ensuremath{\phi}}_{0}$ {Josephson} junctions in twisted bilayer
  graphene},\ }\href {https://doi.org/10.1103/PhysRevResearch.5.L032033}
  {\bibfield  {journal} {\bibinfo  {journal} {Phys. Rev. Res.}\ }\textbf
  {\bibinfo {volume} {5}},\ \bibinfo {pages} {L032033} (\bibinfo {year}
  {2023})}\BibitemShut {NoStop}%
\bibitem [{\citenamefont {Sardinero}\ \emph {et~al.}(2024)\citenamefont
  {Sardinero}, \citenamefont {Seoane~Souto},\ and\ \citenamefont
  {Burset}}]{Sardinero2024}%
  \BibitemOpen
  \bibfield  {author} {\bibinfo {author} {\bibfnamefont {I.}~\bibnamefont
  {Sardinero}}, \bibinfo {author} {\bibfnamefont {R.}~\bibnamefont
  {Seoane~Souto}},\ and\ \bibinfo {author} {\bibfnamefont {P.}~\bibnamefont
  {Burset}},\ }\bibfield  {title} {\bibinfo {title} {Topological
  superconductivity in a magnetic-texture coupled {Josephson} junction},\
  }\href {https://doi.org/10.1103/PhysRevB.110.L060505} {\bibfield  {journal}
  {\bibinfo  {journal} {Phys. Rev. B}\ }\textbf {\bibinfo {volume} {110}},\
  \bibinfo {pages} {L060505} (\bibinfo {year} {2024})}\BibitemShut {NoStop}%
\bibitem [{\citenamefont {Seoane~Souto}\ \emph {et~al.}(2024)\citenamefont
  {Seoane~Souto}, \citenamefont {Kuzmanovski}, \citenamefont {Sardinero},
  \citenamefont {Burset},\ and\ \citenamefont {Balatsky}}]{SeoaneSouto2024Oct}%
  \BibitemOpen
  \bibfield  {author} {\bibinfo {author} {\bibfnamefont {R.}~\bibnamefont
  {Seoane~Souto}}, \bibinfo {author} {\bibfnamefont {D.}~\bibnamefont
  {Kuzmanovski}}, \bibinfo {author} {\bibfnamefont {I.}~\bibnamefont
  {Sardinero}}, \bibinfo {author} {\bibfnamefont {P.}~\bibnamefont {Burset}},\
  and\ \bibinfo {author} {\bibfnamefont {A.~V.}\ \bibnamefont {Balatsky}},\
  }\bibfield  {title} {\bibinfo {title} {P-wave pairing near a spin-split
  {Josephson} junction},\ }\href {https://doi.org/10.1007/s10909-024-03176-0}
  {\bibfield  {journal} {\bibinfo  {journal} {J. Low Temp. Phys.}\ }\textbf
  {\bibinfo {volume} {217}},\ \bibinfo {pages} {106} (\bibinfo {year}
  {2024})}\BibitemShut {NoStop}%
\bibitem [{\citenamefont {Fukaya}\ \emph {et~al.}(2025)\citenamefont {Fukaya},
  \citenamefont {Lu}, \citenamefont {Yada}, \citenamefont {Tanaka},\ and\
  \citenamefont {Cayao}}]{fukayaCayaoReview2025}%
  \BibitemOpen
  \bibfield  {author} {\bibinfo {author} {\bibfnamefont {Y.}~\bibnamefont
  {Fukaya}}, \bibinfo {author} {\bibfnamefont {B.}~\bibnamefont {Lu}}, \bibinfo
  {author} {\bibfnamefont {K.}~\bibnamefont {Yada}}, \bibinfo {author}
  {\bibfnamefont {Y.}~\bibnamefont {Tanaka}},\ and\ \bibinfo {author}
  {\bibfnamefont {J.}~\bibnamefont {Cayao}},\ }\bibfield  {title} {\bibinfo
  {title} {Superconducting phenomena in systems with unconventional magnets},\
  }\href {https://arxiv.org/abs/2502.15400} {\bibfield  {journal} {\bibinfo
  {journal} {arXiv:2502.15400}\ } (\bibinfo {year} {2025})}\BibitemShut
  {NoStop}%
\bibitem [{\citenamefont {Cayao}\ \emph {et~al.}(2020)\citenamefont {Cayao},
  \citenamefont {Triola},\ and\ \citenamefont {Black-Schaffer}}]{Cayao2020odd}%
  \BibitemOpen
  \bibfield  {author} {\bibinfo {author} {\bibfnamefont {J.}~\bibnamefont
  {Cayao}}, \bibinfo {author} {\bibfnamefont {C.}~\bibnamefont {Triola}},\ and\
  \bibinfo {author} {\bibfnamefont {A.~M.}\ \bibnamefont {Black-Schaffer}},\
  }\bibfield  {title} {\bibinfo {title} {Odd-frequency superconducting pairing
  in one-dimensional systems},\ }\href
  {https://doi.org/10.1140/epjst/e2019-900168-0} {\bibfield  {journal}
  {\bibinfo  {journal} {Eur. Phys. J. Special Topics}\ }\textbf {\bibinfo
  {volume} {229}},\ \bibinfo {pages} {545} (\bibinfo {year}
  {2020})}\BibitemShut {NoStop}%
\bibitem [{\citenamefont {Beenakker}(1992)}]{Beenakker:92}%
  \BibitemOpen
  \bibfield  {author} {\bibinfo {author} {\bibfnamefont {C.~W.~J.}\
  \bibnamefont {Beenakker}},\ }\bibfield  {title} {\bibinfo {title} {Three
  {\textquotedblleft}universal{\textquotedblright} mesoscopic {Josephson}
  effects},\ }in\ \href {https://doi.org/10.1007/978-3-642-84818-6_22} {\emph
  {\bibinfo {booktitle} {{Transport Phenomena in Mesoscopic Systems}}}}\
  (\bibinfo  {publisher} {Springer},\ \bibinfo {address} {Berlin, Germany},\
  \bibinfo {year} {1992})\ pp.\ \bibinfo {pages} {235--253}\BibitemShut
  {NoStop}%
\bibitem [{\citenamefont {Mizushima}\ and\ \citenamefont
  {Machida}(2018)}]{mizushima2018multifaceted}%
  \BibitemOpen
  \bibfield  {author} {\bibinfo {author} {\bibfnamefont {T.}~\bibnamefont
  {Mizushima}}\ and\ \bibinfo {author} {\bibfnamefont {K.}~\bibnamefont
  {Machida}},\ }\bibfield  {title} {\bibinfo {title} {Multifaceted properties
  of {A}ndreev bound states: interplay of symmetry and topology},\ }\href
  {https://doi.org/10.1098/rsta.2015.0355} {\bibfield  {journal} {\bibinfo
  {journal} {Philos. Trans. R. Soc. A}\ }\textbf {\bibinfo {volume} {376}},\
  \bibinfo {pages} {20150355} (\bibinfo {year} {2018})}\BibitemShut {NoStop}%
\bibitem [{\citenamefont {Tanaka}\ \emph {et~al.}(2024)\citenamefont {Tanaka},
  \citenamefont {Tamura},\ and\ \citenamefont {Cayao}}]{tanaka2024theory}%
  \BibitemOpen
  \bibfield  {author} {\bibinfo {author} {\bibfnamefont {Y.}~\bibnamefont
  {Tanaka}}, \bibinfo {author} {\bibfnamefont {S.}~\bibnamefont {Tamura}},\
  and\ \bibinfo {author} {\bibfnamefont {J.}~\bibnamefont {Cayao}},\ }\bibfield
   {title} {\bibinfo {title} {Theory of {Majorana} zero modes in unconventional
  superconductors},\ }\href {https://doi.org/10.1093/ptep/ptae065} {\bibfield
  {journal} {\bibinfo  {journal} {Prog. Theor. Exp. Phys.}\ }\textbf {\bibinfo
  {volume} {2024}},\ \bibinfo {pages} {08C105} (\bibinfo {year}
  {2024})}\BibitemShut {NoStop}%
\bibitem [{\citenamefont {Agra\"{\i}t}\ \emph {et~al.}(2003)\citenamefont
  {Agra\"{\i}t}, \citenamefont {Yeyati},\ and\ \citenamefont {van
  Ruitenbeek}}]{AGRAIT200381}%
  \BibitemOpen
  \bibfield  {author} {\bibinfo {author} {\bibfnamefont {N.}~\bibnamefont
  {Agra\"{\i}t}}, \bibinfo {author} {\bibfnamefont {A.~L.}\ \bibnamefont
  {Yeyati}},\ and\ \bibinfo {author} {\bibfnamefont {J.~M.}\ \bibnamefont {van
  Ruitenbeek}},\ }\bibfield  {title} {\bibinfo {title} {{Quantum properties of
  atomic-sized conductors}},\ }\href
  {https://doi.org/10.1016/S0370-1573(02)00633-6} {\bibfield  {journal}
  {\bibinfo  {journal} {Phys. Rep.}\ }\textbf {\bibinfo {volume} {377}},\
  \bibinfo {pages} {81} (\bibinfo {year} {2003})}\BibitemShut {NoStop}%
\bibitem [{\citenamefont {Martín-Rodero}\ and\ \citenamefont
  {Levy~Yeyati}(2011)}]{Martin_Rodero_2011}%
  \BibitemOpen
  \bibfield  {author} {\bibinfo {author} {\bibfnamefont {A.}~\bibnamefont
  {Martín-Rodero}}\ and\ \bibinfo {author} {\bibfnamefont {A.}~\bibnamefont
  {Levy~Yeyati}},\ }\bibfield  {title} {\bibinfo {title} {Josephson and
  {Andreev} transport through quantum dots},\ }\href
  {http://dx.doi.org/10.1080/00018732.2011.624266} {\bibfield  {journal}
  {\bibinfo  {journal} {Advances in Physics}\ }\textbf {\bibinfo {volume}
  {60}},\ \bibinfo {pages} {899–958} (\bibinfo {year} {2011})}\BibitemShut
  {NoStop}%
\bibitem [{\citenamefont {San-Jose}\ \emph {et~al.}(2012)\citenamefont
  {San-Jose}, \citenamefont {Prada},\ and\ \citenamefont
  {Aguado}}]{PhysRevLett.108.257001}%
  \BibitemOpen
  \bibfield  {author} {\bibinfo {author} {\bibfnamefont {P.}~\bibnamefont
  {San-Jose}}, \bibinfo {author} {\bibfnamefont {E.}~\bibnamefont {Prada}},\
  and\ \bibinfo {author} {\bibfnamefont {R.}~\bibnamefont {Aguado}},\
  }\bibfield  {title} {\bibinfo {title} {{AC} {Josephson} effect in
  finite-length nanowire junctions with {Majorana} modes},\ }\href
  {https://doi.org/10.1103/PhysRevLett.108.257001} {\bibfield  {journal}
  {\bibinfo  {journal} {Phys. Rev. Lett.}\ }\textbf {\bibinfo {volume} {108}},\
  \bibinfo {pages} {257001} (\bibinfo {year} {2012})}\BibitemShut {NoStop}%
\bibitem [{\citenamefont {Zazunov}\ \emph {et~al.}(2017)\citenamefont
  {Zazunov}, \citenamefont {Egger}, \citenamefont {Alvarado},\ and\
  \citenamefont {Yeyati}}]{PhysRevB.96.024516}%
  \BibitemOpen
  \bibfield  {author} {\bibinfo {author} {\bibfnamefont {A.}~\bibnamefont
  {Zazunov}}, \bibinfo {author} {\bibfnamefont {R.}~\bibnamefont {Egger}},
  \bibinfo {author} {\bibfnamefont {M.}~\bibnamefont {Alvarado}},\ and\
  \bibinfo {author} {\bibfnamefont {A.~L.}\ \bibnamefont {Yeyati}},\ }\bibfield
   {title} {\bibinfo {title} {Josephson effect in multiterminal topological
  junctions},\ }\href {https://doi.org/10.1103/PhysRevB.96.024516} {\bibfield
  {journal} {\bibinfo  {journal} {Phys. Rev. B}\ }\textbf {\bibinfo {volume}
  {96}},\ \bibinfo {pages} {024516} (\bibinfo {year} {2017})}\BibitemShut
  {NoStop}%
\bibitem [{\citenamefont {Cayao}\ and\ \citenamefont
  {Black-Schaffer}(2018)}]{cayao2018finite}%
  \BibitemOpen
  \bibfield  {author} {\bibinfo {author} {\bibfnamefont {J.}~\bibnamefont
  {Cayao}}\ and\ \bibinfo {author} {\bibfnamefont {A.~M.}\ \bibnamefont
  {Black-Schaffer}},\ }\bibfield  {title} {\bibinfo {title} {{Finite length
  effect on supercurrents between trivial and topological superconductors}},\
  }\href {https://doi.org/10.1140/epjst/e2018-800101-0} {\bibfield  {journal}
  {\bibinfo  {journal} {Eur. Phys. J. Spec. Top.}\ }\textbf {\bibinfo {volume}
  {227}},\ \bibinfo {pages} {1387} (\bibinfo {year} {2018})}\BibitemShut
  {NoStop}%
\bibitem [{\citenamefont {Cayao}\ \emph {et~al.}(2018)\citenamefont {Cayao},
  \citenamefont {Black-Schaffer}, \citenamefont {Prada},\ and\ \citenamefont
  {Aguado}}]{cayao2018andreev}%
  \BibitemOpen
  \bibfield  {author} {\bibinfo {author} {\bibfnamefont {J.}~\bibnamefont
  {Cayao}}, \bibinfo {author} {\bibfnamefont {A.~M.}\ \bibnamefont
  {Black-Schaffer}}, \bibinfo {author} {\bibfnamefont {E.}~\bibnamefont
  {Prada}},\ and\ \bibinfo {author} {\bibfnamefont {R.}~\bibnamefont
  {Aguado}},\ }\bibfield  {title} {\bibinfo {title} {Andreev spectrum and
  supercurrents in nanowire-based {SNS} junctions containing {Majorana} bound
  states},\ }\href {https://www.beilstein-journals.org/bjnano/articles/9/127}
  {\bibfield  {journal} {\bibinfo  {journal} {Beilstein J. Nanotechnol.}\
  }\textbf {\bibinfo {volume} {9}},\ \bibinfo {pages} {1339} (\bibinfo {year}
  {2018})}\BibitemShut {NoStop}%
\bibitem [{\citenamefont {Beenakker}\ and\ \citenamefont
  {Vakhtel}(2023)}]{PhysRevB.108.075425}%
  \BibitemOpen
  \bibfield  {author} {\bibinfo {author} {\bibfnamefont {C.~W.~J.}\
  \bibnamefont {Beenakker}}\ and\ \bibinfo {author} {\bibfnamefont
  {T.}~\bibnamefont {Vakhtel}},\ }\bibfield  {title} {\bibinfo {title}
  {Phase-shifted {Andreev} levels in an altermagnet {Josephson} junction},\
  }\href {https://doi.org/10.1103/PhysRevB.108.075425} {\bibfield  {journal}
  {\bibinfo  {journal} {Phys. Rev. B}\ }\textbf {\bibinfo {volume} {108}},\
  \bibinfo {pages} {075425} (\bibinfo {year} {2023})}\BibitemShut {NoStop}%
\bibitem [{\citenamefont {Awoga}\ \emph {et~al.}(2019)\citenamefont {Awoga},
  \citenamefont {Cayao},\ and\ \citenamefont
  {Black-Schaffer}}]{PhysRevLett.123.117001}%
  \BibitemOpen
  \bibfield  {author} {\bibinfo {author} {\bibfnamefont {O.~A.}\ \bibnamefont
  {Awoga}}, \bibinfo {author} {\bibfnamefont {J.}~\bibnamefont {Cayao}},\ and\
  \bibinfo {author} {\bibfnamefont {A.~M.}\ \bibnamefont {Black-Schaffer}},\
  }\bibfield  {title} {\bibinfo {title} {Supercurrent detection of
  topologically trivial zero-energy states in nanowire junctions},\ }\href
  {https://doi.org/10.1103/PhysRevLett.123.117001} {\bibfield  {journal}
  {\bibinfo  {journal} {Phys. Rev. Lett.}\ }\textbf {\bibinfo {volume} {123}},\
  \bibinfo {pages} {117001} (\bibinfo {year} {2019})}\BibitemShut {NoStop}%
\bibitem [{\citenamefont {Cayao}\ and\ \citenamefont
  {Black-Schaffer}(2021)}]{PhysRevB.104.L020501}%
  \BibitemOpen
  \bibfield  {author} {\bibinfo {author} {\bibfnamefont {J.}~\bibnamefont
  {Cayao}}\ and\ \bibinfo {author} {\bibfnamefont {A.~M.}\ \bibnamefont
  {Black-Schaffer}},\ }\bibfield  {title} {\bibinfo {title} {Distinguishing
  trivial and topological zero-energy states in long nanowire junctions},\
  }\href {https://doi.org/10.1103/PhysRevB.104.L020501} {\bibfield  {journal}
  {\bibinfo  {journal} {Phys. Rev. B}\ }\textbf {\bibinfo {volume} {104}},\
  \bibinfo {pages} {L020501} (\bibinfo {year} {2021})}\BibitemShut {NoStop}%
\bibitem [{\citenamefont {Zhao}\ \emph {et~al.}(2025)\citenamefont {Zhao},
  \citenamefont {Fukaya}, \citenamefont {Burset}, \citenamefont {Cayao},
  \citenamefont {Tanaka},\ and\ \citenamefont {Lu}}]{Zhao2025}%
  \BibitemOpen
  \bibfield  {author} {\bibinfo {author} {\bibfnamefont {W.}~\bibnamefont
  {Zhao}}, \bibinfo {author} {\bibfnamefont {Y.}~\bibnamefont {Fukaya}},
  \bibinfo {author} {\bibfnamefont {P.}~\bibnamefont {Burset}}, \bibinfo
  {author} {\bibfnamefont {J.}~\bibnamefont {Cayao}}, \bibinfo {author}
  {\bibfnamefont {Y.}~\bibnamefont {Tanaka}},\ and\ \bibinfo {author}
  {\bibfnamefont {B.}~\bibnamefont {Lu}},\ }\bibfield  {title} {\bibinfo
  {title} {Orientation-dependent transport in junctions formed by $d$-wave
  altermagnets and $d$-wave superconductors},\ }\href
  {https://doi.org/10.1103/PhysRevB.111.184515} {\bibfield  {journal} {\bibinfo
   {journal} {Phys. Rev. B}\ }\textbf {\bibinfo {volume} {111}},\ \bibinfo
  {pages} {184515} (\bibinfo {year} {2025})}\BibitemShut {NoStop}%
\bibitem [{\citenamefont {Cayao}\ \emph
  {et~al.}(2022{\natexlab{a}})\citenamefont {Cayao}, \citenamefont {Dutta},
  \citenamefont {Burset},\ and\ \citenamefont
  {Black-Schaffer}}]{PhysRevB.106.L100502}%
  \BibitemOpen
  \bibfield  {author} {\bibinfo {author} {\bibfnamefont {J.}~\bibnamefont
  {Cayao}}, \bibinfo {author} {\bibfnamefont {P.}~\bibnamefont {Dutta}},
  \bibinfo {author} {\bibfnamefont {P.}~\bibnamefont {Burset}},\ and\ \bibinfo
  {author} {\bibfnamefont {A.~M.}\ \bibnamefont {Black-Schaffer}},\ }\bibfield
  {title} {\bibinfo {title} {Phase-tunable electron transport assisted by
  odd-frequency {Cooper} pairs in topological {Josephson} junctions},\ }\href
  {https://doi.org/10.1103/PhysRevB.106.L100502} {\bibfield  {journal}
  {\bibinfo  {journal} {Phys. Rev. B}\ }\textbf {\bibinfo {volume} {106}},\
  \bibinfo {pages} {L100502} (\bibinfo {year}
  {2022}{\natexlab{a}})}\BibitemShut {NoStop}%
\bibitem [{\citenamefont {Yang}\ \emph {et~al.}(2023)\citenamefont {Yang},
  \citenamefont {Burset},\ and\ \citenamefont {Lu}}]{Yang_2023}%
  \BibitemOpen
  \bibfield  {author} {\bibinfo {author} {\bibfnamefont {X.}~\bibnamefont
  {Yang}}, \bibinfo {author} {\bibfnamefont {P.}~\bibnamefont {Burset}},\ and\
  \bibinfo {author} {\bibfnamefont {B.}~\bibnamefont {Lu}},\ }\bibfield
  {title} {\bibinfo {title} {Phase-tunable multiple {Andreev} reflections in a
  quantum spin {Hall} strip},\ }\href
  {https://doi.org/10.1088/1361-6668/ace2f0} {\bibfield  {journal} {\bibinfo
  {journal} {Superconductor Science and Technology}\ }\textbf {\bibinfo
  {volume} {36}},\ \bibinfo {pages} {085012} (\bibinfo {year}
  {2023})}\BibitemShut {NoStop}%
\bibitem [{\citenamefont {Lu}\ \emph {et~al.}(2020)\citenamefont {Lu},
  \citenamefont {Burset},\ and\ \citenamefont {Tanaka}}]{Lu2020Jan}%
  \BibitemOpen
  \bibfield  {author} {\bibinfo {author} {\bibfnamefont {B.}~\bibnamefont
  {Lu}}, \bibinfo {author} {\bibfnamefont {P.}~\bibnamefont {Burset}},\ and\
  \bibinfo {author} {\bibfnamefont {Y.}~\bibnamefont {Tanaka}},\ }\bibfield
  {title} {\bibinfo {title} {Spin-polarized multiple {Andreev} reflections in
  spin-split superconductors},\ }\href
  {https://doi.org/10.1103/PhysRevB.101.020502} {\bibfield  {journal} {\bibinfo
   {journal} {Phys. Rev. B}\ }\textbf {\bibinfo {volume} {101}},\ \bibinfo
  {pages} {020502} (\bibinfo {year} {2020})}\BibitemShut {NoStop}%
\bibitem [{\citenamefont {Cayao}\ \emph
  {et~al.}(2022{\natexlab{b}})\citenamefont {Cayao}, \citenamefont {Dutta},
  \citenamefont {Burset},\ and\ \citenamefont {Black-Schaffer}}]{Cayao2022Sep}%
  \BibitemOpen
  \bibfield  {author} {\bibinfo {author} {\bibfnamefont {J.}~\bibnamefont
  {Cayao}}, \bibinfo {author} {\bibfnamefont {P.}~\bibnamefont {Dutta}},
  \bibinfo {author} {\bibfnamefont {P.}~\bibnamefont {Burset}},\ and\ \bibinfo
  {author} {\bibfnamefont {A.~M.}\ \bibnamefont {Black-Schaffer}},\ }\bibfield
  {title} {\bibinfo {title} {Phase-tunable electron transport assisted by
  odd-frequency {Cooper} pairs in topological {Josephson} junctions},\ }\href
  {https://doi.org/10.1103/PhysRevB.106.L100502} {\bibfield  {journal}
  {\bibinfo  {journal} {Phys. Rev. B}\ }\textbf {\bibinfo {volume} {106}},\
  \bibinfo {pages} {L100502} (\bibinfo {year}
  {2022}{\natexlab{b}})}\BibitemShut {NoStop}%
\bibitem [{\citenamefont {Cayao}\ \emph {et~al.}(2024)\citenamefont {Cayao},
  \citenamefont {Burset},\ and\ \citenamefont {Tanaka}}]{Cayao2024May}%
  \BibitemOpen
  \bibfield  {author} {\bibinfo {author} {\bibfnamefont {J.}~\bibnamefont
  {Cayao}}, \bibinfo {author} {\bibfnamefont {P.}~\bibnamefont {Burset}},\ and\
  \bibinfo {author} {\bibfnamefont {Y.}~\bibnamefont {Tanaka}},\ }\bibfield
  {title} {\bibinfo {title} {Controllable odd-frequency {Cooper} pairs in
  multisuperconductor {Josephson} junctions},\ }\href
  {https://doi.org/10.1103/PhysRevB.109.205406} {\bibfield  {journal} {\bibinfo
   {journal} {Phys. Rev. B}\ }\textbf {\bibinfo {volume} {109}},\ \bibinfo
  {pages} {205406} (\bibinfo {year} {2024})}\BibitemShut {NoStop}%
\bibitem [{\citenamefont {Cayao}\ and\ \citenamefont
  {Sato}(2024{\natexlab{a}})}]{PhysRevB.110.L201403}%
  \BibitemOpen
  \bibfield  {author} {\bibinfo {author} {\bibfnamefont {J.}~\bibnamefont
  {Cayao}}\ and\ \bibinfo {author} {\bibfnamefont {M.}~\bibnamefont {Sato}},\
  }\bibfield  {title} {\bibinfo {title} {Non-{Hermitian} phase-biased josephson
  junctions},\ }\href {https://doi.org/10.1103/PhysRevB.110.L201403} {\bibfield
   {journal} {\bibinfo  {journal} {Phys. Rev. B}\ }\textbf {\bibinfo {volume}
  {110}},\ \bibinfo {pages} {L201403} (\bibinfo {year}
  {2024}{\natexlab{a}})}\BibitemShut {NoStop}%
\bibitem [{\citenamefont {Nesterov}\ \emph {et~al.}(2016)\citenamefont
  {Nesterov}, \citenamefont {Houzet},\ and\ \citenamefont
  {Meyer}}]{PhysRevB.93.174502}%
  \BibitemOpen
  \bibfield  {author} {\bibinfo {author} {\bibfnamefont {K.~N.}\ \bibnamefont
  {Nesterov}}, \bibinfo {author} {\bibfnamefont {M.}~\bibnamefont {Houzet}},\
  and\ \bibinfo {author} {\bibfnamefont {J.~S.}\ \bibnamefont {Meyer}},\
  }\bibfield  {title} {\bibinfo {title} {Anomalous {Josephson} effect in
  semiconducting nanowires as a signature of the topologically nontrivial
  phase},\ }\href {https://doi.org/10.1103/PhysRevB.93.174502} {\bibfield
  {journal} {\bibinfo  {journal} {Phys. Rev. B}\ }\textbf {\bibinfo {volume}
  {93}},\ \bibinfo {pages} {174502} (\bibinfo {year} {2016})}\BibitemShut
  {NoStop}%
\bibitem [{\citenamefont {Ruiz}\ \emph {et~al.}(2022)\citenamefont {Ruiz},
  \citenamefont {Rampp}, \citenamefont {Aligia}, \citenamefont {Schmalian},\
  and\ \citenamefont {Arrachea}}]{PhysRevB.106.195415}%
  \BibitemOpen
  \bibfield  {author} {\bibinfo {author} {\bibfnamefont {G.~F.~R.}\
  \bibnamefont {Ruiz}}, \bibinfo {author} {\bibfnamefont {M.~A.}\ \bibnamefont
  {Rampp}}, \bibinfo {author} {\bibfnamefont {A.~A.}\ \bibnamefont {Aligia}},
  \bibinfo {author} {\bibfnamefont {J.}~\bibnamefont {Schmalian}},\ and\
  \bibinfo {author} {\bibfnamefont {L.}~\bibnamefont {Arrachea}},\ }\bibfield
  {title} {\bibinfo {title} {Josephson junctions of two-dimensional
  time-reversal invariant superconductors: Signatures of the topological
  phase},\ }\href {https://doi.org/10.1103/PhysRevB.106.195415} {\bibfield
  {journal} {\bibinfo  {journal} {Phys. Rev. B}\ }\textbf {\bibinfo {volume}
  {106}},\ \bibinfo {pages} {195415} (\bibinfo {year} {2022})}\BibitemShut
  {NoStop}%
\bibitem [{\citenamefont {Cayao}\ and\ \citenamefont
  {Sato}(2024{\natexlab{b}})}]{PhysRevB.110.235426}%
  \BibitemOpen
  \bibfield  {author} {\bibinfo {author} {\bibfnamefont {J.}~\bibnamefont
  {Cayao}}\ and\ \bibinfo {author} {\bibfnamefont {M.}~\bibnamefont {Sato}},\
  }\bibfield  {title} {\bibinfo {title} {Non-{Hermitian} multiterminal
  phase-biased josephson junctions},\ }\href
  {https://doi.org/10.1103/PhysRevB.110.235426} {\bibfield  {journal} {\bibinfo
   {journal} {Phys. Rev. B}\ }\textbf {\bibinfo {volume} {110}},\ \bibinfo
  {pages} {235426} (\bibinfo {year} {2024}{\natexlab{b}})}\BibitemShut
  {NoStop}%
\bibitem [{\citenamefont {Dolcini}\ \emph {et~al.}(2015)\citenamefont
  {Dolcini}, \citenamefont {Houzet},\ and\ \citenamefont
  {Meyer}}]{PhysRevB.92.035428}%
  \BibitemOpen
  \bibfield  {author} {\bibinfo {author} {\bibfnamefont {F.}~\bibnamefont
  {Dolcini}}, \bibinfo {author} {\bibfnamefont {M.}~\bibnamefont {Houzet}},\
  and\ \bibinfo {author} {\bibfnamefont {J.~S.}\ \bibnamefont {Meyer}},\
  }\bibfield  {title} {\bibinfo {title} {Topological {Josephson}
  ${\ensuremath{\phi}}_{0}$ junctions},\ }\href
  {https://doi.org/10.1103/PhysRevB.92.035428} {\bibfield  {journal} {\bibinfo
  {journal} {Phys. Rev. B}\ }\textbf {\bibinfo {volume} {92}},\ \bibinfo
  {pages} {035428} (\bibinfo {year} {2015})}\BibitemShut {NoStop}%
\bibitem [{\citenamefont {Goffman}\ \emph {et~al.}(2000)\citenamefont
  {Goffman}, \citenamefont {Cron}, \citenamefont {Levy~Yeyati}, \citenamefont
  {Joyez}, \citenamefont {Devoret}, \citenamefont {Esteve},\ and\ \citenamefont
  {Urbina}}]{goffman2000supercurrent}%
  \BibitemOpen
  \bibfield  {author} {\bibinfo {author} {\bibfnamefont {M.~F.}\ \bibnamefont
  {Goffman}}, \bibinfo {author} {\bibfnamefont {R.}~\bibnamefont {Cron}},
  \bibinfo {author} {\bibfnamefont {A.}~\bibnamefont {Levy~Yeyati}}, \bibinfo
  {author} {\bibfnamefont {P.}~\bibnamefont {Joyez}}, \bibinfo {author}
  {\bibfnamefont {M.~H.}\ \bibnamefont {Devoret}}, \bibinfo {author}
  {\bibfnamefont {D.}~\bibnamefont {Esteve}},\ and\ \bibinfo {author}
  {\bibfnamefont {C.}~\bibnamefont {Urbina}},\ }\bibfield  {title} {\bibinfo
  {title} {Supercurrent in atomic point contacts and {Andreev} states},\ }\href
  {https://doi.org/10.1103/PhysRevLett.85.170} {\bibfield  {journal} {\bibinfo
  {journal} {Phys. Rev. Lett.}\ }\textbf {\bibinfo {volume} {85}},\ \bibinfo
  {pages} {170} (\bibinfo {year} {2000})}\BibitemShut {NoStop}%
\bibitem [{\citenamefont {Pillet}\ \emph {et~al.}(2010)\citenamefont {Pillet},
  \citenamefont {Quay}, \citenamefont {Morfin}, \citenamefont {Bena},
  \citenamefont {Yeyati},\ and\ \citenamefont {Joyez}}]{pillet2010andreev}%
  \BibitemOpen
  \bibfield  {author} {\bibinfo {author} {\bibfnamefont {J.-D.}\ \bibnamefont
  {Pillet}}, \bibinfo {author} {\bibfnamefont {C.~H.~L.}\ \bibnamefont {Quay}},
  \bibinfo {author} {\bibfnamefont {P.}~\bibnamefont {Morfin}}, \bibinfo
  {author} {\bibfnamefont {C.}~\bibnamefont {Bena}}, \bibinfo {author}
  {\bibfnamefont {A.~L.}\ \bibnamefont {Yeyati}},\ and\ \bibinfo {author}
  {\bibfnamefont {P.}~\bibnamefont {Joyez}},\ }\bibfield  {title} {\bibinfo
  {title} {{Andreev bound states in supercurrent-carrying carbon nanotubes
  revealed}},\ }\href {https://doi.org/10.1038/nphys1811} {\bibfield  {journal}
  {\bibinfo  {journal} {Nat. Phys.}\ }\textbf {\bibinfo {volume} {6}},\
  \bibinfo {pages} {965} (\bibinfo {year} {2010})}\BibitemShut {NoStop}%
\bibitem [{\citenamefont {Spanton}\ \emph {et~al.}(2017)\citenamefont
  {Spanton}, \citenamefont {Deng}, \citenamefont
  {Vaitiek{\ifmmode\dot{e}\else\.{e}\fi}nas}, \citenamefont {Krogstrup},
  \citenamefont {Nyg{\aa}rd}, \citenamefont {Marcus},\ and\ \citenamefont
  {Moler}}]{spanton2017current}%
  \BibitemOpen
  \bibfield  {author} {\bibinfo {author} {\bibfnamefont {E.~M.}\ \bibnamefont
  {Spanton}}, \bibinfo {author} {\bibfnamefont {M.}~\bibnamefont {Deng}},
  \bibinfo {author} {\bibfnamefont {S.}~\bibnamefont
  {Vaitiek{\ifmmode\dot{e}\else\.{e}\fi}nas}}, \bibinfo {author} {\bibfnamefont
  {P.}~\bibnamefont {Krogstrup}}, \bibinfo {author} {\bibfnamefont
  {J.}~\bibnamefont {Nyg{\aa}rd}}, \bibinfo {author} {\bibfnamefont {C.~M.}\
  \bibnamefont {Marcus}},\ and\ \bibinfo {author} {\bibfnamefont {K.~A.}\
  \bibnamefont {Moler}},\ }\bibfield  {title} {\bibinfo {title}
  {{Current{\textendash}phase relations of few-mode InAs nanowire Josephson
  junctions}},\ }\href {https://doi.org/10.1038/nphys4224} {\bibfield
  {journal} {\bibinfo  {journal} {Nat. Phys.}\ }\textbf {\bibinfo {volume}
  {13}},\ \bibinfo {pages} {1177} (\bibinfo {year} {2017})}\BibitemShut
  {NoStop}%
\bibitem [{\citenamefont {Nichele}\ \emph {et~al.}(2020)\citenamefont
  {Nichele}, \citenamefont {Portol\'es}, \citenamefont {Fornieri},
  \citenamefont {Whiticar}, \citenamefont {Drachmann}, \citenamefont {Gronin},
  \citenamefont {Wang}, \citenamefont {Gardner}, \citenamefont {Thomas},
  \citenamefont {Hatke}, \citenamefont {Manfra},\ and\ \citenamefont
  {Marcus}}]{PhysRevLett.124.226801}%
  \BibitemOpen
  \bibfield  {author} {\bibinfo {author} {\bibfnamefont {F.}~\bibnamefont
  {Nichele}}, \bibinfo {author} {\bibfnamefont {E.}~\bibnamefont {Portol\'es}},
  \bibinfo {author} {\bibfnamefont {A.}~\bibnamefont {Fornieri}}, \bibinfo
  {author} {\bibfnamefont {A.~M.}\ \bibnamefont {Whiticar}}, \bibinfo {author}
  {\bibfnamefont {A.~C.~C.}\ \bibnamefont {Drachmann}}, \bibinfo {author}
  {\bibfnamefont {S.}~\bibnamefont {Gronin}}, \bibinfo {author} {\bibfnamefont
  {T.}~\bibnamefont {Wang}}, \bibinfo {author} {\bibfnamefont {G.~C.}\
  \bibnamefont {Gardner}}, \bibinfo {author} {\bibfnamefont {C.}~\bibnamefont
  {Thomas}}, \bibinfo {author} {\bibfnamefont {A.~T.}\ \bibnamefont {Hatke}},
  \bibinfo {author} {\bibfnamefont {M.~J.}\ \bibnamefont {Manfra}},\ and\
  \bibinfo {author} {\bibfnamefont {C.~M.}\ \bibnamefont {Marcus}},\ }\bibfield
   {title} {\bibinfo {title} {Relating {Andreev} bound states and supercurrents
  in hybrid {Josephson} junctions},\ }\href
  {https://doi.org/10.1103/PhysRevLett.124.226801} {\bibfield  {journal}
  {\bibinfo  {journal} {Phys. Rev. Lett.}\ }\textbf {\bibinfo {volume} {124}},\
  \bibinfo {pages} {226801} (\bibinfo {year} {2020})}\BibitemShut {NoStop}%
\bibitem [{\citenamefont {Razmadze}\ \emph {et~al.}(2020)\citenamefont
  {Razmadze}, \citenamefont {O'Farrell}, \citenamefont {Krogstrup},\ and\
  \citenamefont {Marcus}}]{PhysRevLett.125.116803}%
  \BibitemOpen
  \bibfield  {author} {\bibinfo {author} {\bibfnamefont {D.}~\bibnamefont
  {Razmadze}}, \bibinfo {author} {\bibfnamefont {E.~C.~T.}\ \bibnamefont
  {O'Farrell}}, \bibinfo {author} {\bibfnamefont {P.}~\bibnamefont
  {Krogstrup}},\ and\ \bibinfo {author} {\bibfnamefont {C.~M.}\ \bibnamefont
  {Marcus}},\ }\bibfield  {title} {\bibinfo {title} {Quantum dot parity effects
  in trivial and topological {Josephson} junctions},\ }\href
  {https://doi.org/10.1103/PhysRevLett.125.116803} {\bibfield  {journal}
  {\bibinfo  {journal} {Phys. Rev. Lett.}\ }\textbf {\bibinfo {volume} {125}},\
  \bibinfo {pages} {116803} (\bibinfo {year} {2020})}\BibitemShut {NoStop}%
\bibitem [{\citenamefont {Janvier}\ \emph {et~al.}(2015)\citenamefont
  {Janvier}, \citenamefont {Tosi}, \citenamefont {Bretheau}, \citenamefont
  {Girit}, \citenamefont {Stern}, \citenamefont {Bertet}, \citenamefont
  {Joyez}, \citenamefont {Vion}, \citenamefont {Esteve}, \citenamefont
  {Goffman}, \citenamefont {Pothier},\ and\ \citenamefont
  {Urbina}}]{Janvier2015}%
  \BibitemOpen
  \bibfield  {author} {\bibinfo {author} {\bibfnamefont {C.}~\bibnamefont
  {Janvier}}, \bibinfo {author} {\bibfnamefont {L.}~\bibnamefont {Tosi}},
  \bibinfo {author} {\bibfnamefont {L.}~\bibnamefont {Bretheau}}, \bibinfo
  {author} {\bibfnamefont
  {{\ifmmode\mbox{\c{C}}\else\c{C}\fi}.~{\ifmmode\ddot{O}\else\"{O}\fi}.}\
  \bibnamefont {Girit}}, \bibinfo {author} {\bibfnamefont {M.}~\bibnamefont
  {Stern}}, \bibinfo {author} {\bibfnamefont {P.}~\bibnamefont {Bertet}},
  \bibinfo {author} {\bibfnamefont {P.}~\bibnamefont {Joyez}}, \bibinfo
  {author} {\bibfnamefont {D.}~\bibnamefont {Vion}}, \bibinfo {author}
  {\bibfnamefont {D.}~\bibnamefont {Esteve}}, \bibinfo {author} {\bibfnamefont
  {M.~F.}\ \bibnamefont {Goffman}}, \bibinfo {author} {\bibfnamefont
  {H.}~\bibnamefont {Pothier}},\ and\ \bibinfo {author} {\bibfnamefont
  {C.}~\bibnamefont {Urbina}},\ }\bibfield  {title} {\bibinfo {title}
  {{Coherent manipulation of Andreev states in superconducting atomic
  contacts}},\ }\href {https://doi.org/10.1126/science.aab2179} {\bibfield
  {journal} {\bibinfo  {journal} {Science}\ }\textbf {\bibinfo {volume}
  {349}},\ \bibinfo {pages} {1199} (\bibinfo {year} {2015})}\BibitemShut
  {NoStop}%
\bibitem [{\citenamefont {Tosi}\ \emph {et~al.}(2019)\citenamefont {Tosi},
  \citenamefont {Metzger}, \citenamefont {Goffman}, \citenamefont {Urbina},
  \citenamefont {Pothier}, \citenamefont {Park}, \citenamefont {Yeyati},
  \citenamefont {Nyg\aa{}rd},\ and\ \citenamefont {Krogstrup}}]{Tosi2019}%
  \BibitemOpen
  \bibfield  {author} {\bibinfo {author} {\bibfnamefont {L.}~\bibnamefont
  {Tosi}}, \bibinfo {author} {\bibfnamefont {C.}~\bibnamefont {Metzger}},
  \bibinfo {author} {\bibfnamefont {M.~F.}\ \bibnamefont {Goffman}}, \bibinfo
  {author} {\bibfnamefont {C.}~\bibnamefont {Urbina}}, \bibinfo {author}
  {\bibfnamefont {H.}~\bibnamefont {Pothier}}, \bibinfo {author} {\bibfnamefont
  {S.}~\bibnamefont {Park}}, \bibinfo {author} {\bibfnamefont {A.~L.}\
  \bibnamefont {Yeyati}}, \bibinfo {author} {\bibfnamefont {J.}~\bibnamefont
  {Nyg\aa{}rd}},\ and\ \bibinfo {author} {\bibfnamefont {P.}~\bibnamefont
  {Krogstrup}},\ }\bibfield  {title} {\bibinfo {title} {Spin-orbit splitting of
  andreev states revealed by microwave spectroscopy},\ }\href
  {https://doi.org/10.1103/PhysRevX.9.011010} {\bibfield  {journal} {\bibinfo
  {journal} {Phys. Rev. X}\ }\textbf {\bibinfo {volume} {9}},\ \bibinfo {pages}
  {011010} (\bibinfo {year} {2019})}\BibitemShut {NoStop}%
\bibitem [{\citenamefont {Jagannathan}(2021)}]{Jagannathan_RMP_2021}%
  \BibitemOpen
  \bibfield  {author} {\bibinfo {author} {\bibfnamefont {A.}~\bibnamefont
  {Jagannathan}},\ }\bibfield  {title} {\bibinfo {title} {{The Fibonacci
  quasicrystal: Case study of hidden dimensions and multifractality}},\ }\href
  {https://doi.org/10.1103/RevModPhys.93.045001} {\bibfield  {journal}
  {\bibinfo  {journal} {Rev. Mod. Phys.}\ }\textbf {\bibinfo {volume} {93}},\
  \bibinfo {pages} {045001} (\bibinfo {year} {2021})}\BibitemShut {NoStop}%
\bibitem [{\citenamefont {Kamiya}\ \emph {et~al.}(2018)\citenamefont {Kamiya},
  \citenamefont {Takeuchi}, \citenamefont {Kabeya}, \citenamefont {Wada},
  \citenamefont {Ishimasa}, \citenamefont {Ochiai}, \citenamefont {Deguchi},
  \citenamefont {Imura},\ and\ \citenamefont {Sato}}]{Kamiya_NatComm2018}%
  \BibitemOpen
  \bibfield  {author} {\bibinfo {author} {\bibfnamefont {K.}~\bibnamefont
  {Kamiya}}, \bibinfo {author} {\bibfnamefont {T.}~\bibnamefont {Takeuchi}},
  \bibinfo {author} {\bibfnamefont {N.}~\bibnamefont {Kabeya}}, \bibinfo
  {author} {\bibfnamefont {N.}~\bibnamefont {Wada}}, \bibinfo {author}
  {\bibfnamefont {T.}~\bibnamefont {Ishimasa}}, \bibinfo {author}
  {\bibfnamefont {A.}~\bibnamefont {Ochiai}}, \bibinfo {author} {\bibfnamefont
  {K.}~\bibnamefont {Deguchi}}, \bibinfo {author} {\bibfnamefont
  {K.}~\bibnamefont {Imura}},\ and\ \bibinfo {author} {\bibfnamefont {N.~K.}\
  \bibnamefont {Sato}},\ }\bibfield  {title} {\bibinfo {title} {{Discovery of
  superconductivity in quasicrystal}},\ }\href
  {https://doi.org/10.1038/s41467-017-02667-x} {\bibfield  {journal} {\bibinfo
  {journal} {Nat. Commun.}\ }\textbf {\bibinfo {volume} {9}},\ \bibinfo {pages}
  {1} (\bibinfo {year} {2018})}\BibitemShut {NoStop}%
\bibitem [{\citenamefont {Wang}\ \emph {et~al.}(2024)\citenamefont {Wang},
  \citenamefont {Rai}, \citenamefont {Matsumura}, \citenamefont {Jagannathan},\
  and\ \citenamefont {Haas}}]{Wang_arxiv_2024}%
  \BibitemOpen
  \bibfield  {author} {\bibinfo {author} {\bibfnamefont {Y.}~\bibnamefont
  {Wang}}, \bibinfo {author} {\bibfnamefont {G.}~\bibnamefont {Rai}}, \bibinfo
  {author} {\bibfnamefont {C.}~\bibnamefont {Matsumura}}, \bibinfo {author}
  {\bibfnamefont {A.}~\bibnamefont {Jagannathan}},\ and\ \bibinfo {author}
  {\bibfnamefont {S.}~\bibnamefont {Haas}},\ }\bibfield  {title} {\bibinfo
  {title} {{Superconductivity in the Fibonacci chain}},\ }\href
  {https://arxiv.org/abs/2403.06157} {\bibfield  {journal} {\bibinfo  {journal}
  {arXiv.2403.06157}\ } (\bibinfo {year} {2024})}\BibitemShut {NoStop}%
\bibitem [{\citenamefont {Sandberg}\ \emph {et~al.}(2024)\citenamefont
  {Sandberg}, \citenamefont {Awoga}, \citenamefont {Black-Schaffer},\ and\
  \citenamefont {Holmvall}}]{Sandberg_PRB_2024}%
  \BibitemOpen
  \bibfield  {author} {\bibinfo {author} {\bibfnamefont {A.}~\bibnamefont
  {Sandberg}}, \bibinfo {author} {\bibfnamefont {O.~A.}\ \bibnamefont {Awoga}},
  \bibinfo {author} {\bibfnamefont {A.~M.}\ \bibnamefont {Black-Schaffer}},\
  and\ \bibinfo {author} {\bibfnamefont {P.}~\bibnamefont {Holmvall}},\
  }\bibfield  {title} {\bibinfo {title} {Josephson effect in a {Fibonacci}
  quasicrystal},\ }\href {https://doi.org/10.1103/PhysRevB.110.104513}
  {\bibfield  {journal} {\bibinfo  {journal} {Phys. Rev. B}\ }\textbf {\bibinfo
  {volume} {110}},\ \bibinfo {pages} {104513} (\bibinfo {year}
  {2024})}\BibitemShut {NoStop}%
\bibitem [{\citenamefont {Kobia\l{}ka}\ \emph {et~al.}(2024)\citenamefont
  {Kobia\l{}ka}, \citenamefont {Awoga}, \citenamefont {Leijnse}, \citenamefont
  {Doma\ifmmode~\acute{n}\else \'{n}\fi{}ski}, \citenamefont {Holmvall},\ and\
  \citenamefont {Black-Schaffer}}]{Kobialka_PRB_2024}%
  \BibitemOpen
  \bibfield  {author} {\bibinfo {author} {\bibfnamefont {A.}~\bibnamefont
  {Kobia\l{}ka}}, \bibinfo {author} {\bibfnamefont {O.~A.}\ \bibnamefont
  {Awoga}}, \bibinfo {author} {\bibfnamefont {M.}~\bibnamefont {Leijnse}},
  \bibinfo {author} {\bibfnamefont {T.}~\bibnamefont
  {Doma\ifmmode~\acute{n}\else \'{n}\fi{}ski}}, \bibinfo {author}
  {\bibfnamefont {P.}~\bibnamefont {Holmvall}},\ and\ \bibinfo {author}
  {\bibfnamefont {A.~M.}\ \bibnamefont {Black-Schaffer}},\ }\bibfield  {title}
  {\bibinfo {title} {Topological superconductivity in {Fibonacci}
  quasicrystals},\ }\href {https://doi.org/10.1103/PhysRevB.110.134508}
  {\bibfield  {journal} {\bibinfo  {journal} {Phys. Rev. B}\ }\textbf {\bibinfo
  {volume} {110}},\ \bibinfo {pages} {134508} (\bibinfo {year}
  {2024})}\BibitemShut {NoStop}%
\bibitem [{\citenamefont {Kraus}\ and\ \citenamefont
  {Zilberberg}(2012)}]{Kraus_PRL2012b}%
  \BibitemOpen
  \bibfield  {author} {\bibinfo {author} {\bibfnamefont {Y.~E.}\ \bibnamefont
  {Kraus}}\ and\ \bibinfo {author} {\bibfnamefont {O.}~\bibnamefont
  {Zilberberg}},\ }\bibfield  {title} {\bibinfo {title} {Topological
  equivalence between the {Fibonacci} quasicrystal and the {Harper} model},\
  }\href {https://doi.org/10.1103/PhysRevLett.109.116404} {\bibfield  {journal}
  {\bibinfo  {journal} {Phys. Rev. Lett.}\ }\textbf {\bibinfo {volume} {109}},\
  \bibinfo {pages} {116404} (\bibinfo {year} {2012})}\BibitemShut {NoStop}%
\bibitem [{SM()}]{SM}%
  \BibitemOpen
  \href@noop {} {}\bibinfo {note} {See Supplemental Material (SM), including
  Refs.~\cite{Cuevas_WS2017, deGennes_WAB_1966, Rai_PRB2019, Wang_arxiv_2024},
  where we provide additional details about (1) how we define the Fibonacci
  sequence and the approximants; (2) the impact of the quasiperiodic potential
  on the junction transmission; (3) how superconducting correlations develop
  inside the Fibonacci gaps; (4) how Fibonacci-Andreev bound states become the
  dominant contribution to the supercurrent; (5) the stability of the Fibonacci
  gaps; and (6) the self-consistency of the superconducting order
  parameter.}\BibitemShut {Stop}%
\bibitem [{\citenamefont {Kraus}\ \emph {et~al.}(2012)\citenamefont {Kraus},
  \citenamefont {Lahini}, \citenamefont {Ringel}, \citenamefont {Verbin},\ and\
  \citenamefont {Zilberberg}}]{Kraus_PRL2012a}%
  \BibitemOpen
  \bibfield  {author} {\bibinfo {author} {\bibfnamefont {Y.~E.}\ \bibnamefont
  {Kraus}}, \bibinfo {author} {\bibfnamefont {Y.}~\bibnamefont {Lahini}},
  \bibinfo {author} {\bibfnamefont {Z.}~\bibnamefont {Ringel}}, \bibinfo
  {author} {\bibfnamefont {M.}~\bibnamefont {Verbin}},\ and\ \bibinfo {author}
  {\bibfnamefont {O.}~\bibnamefont {Zilberberg}},\ }\bibfield  {title}
  {\bibinfo {title} {{Topological States and Adiabatic Pumping in
  Quasicrystals}},\ }\href {https://doi.org/10.1103/PhysRevLett.109.106402}
  {\bibfield  {journal} {\bibinfo  {journal} {Phys. Rev. Lett.}\ }\textbf
  {\bibinfo {volume} {109}},\ \bibinfo {pages} {106402} (\bibinfo {year}
  {2012})}\BibitemShut {NoStop}%
\bibitem [{Note1()}]{Note1}%
  \BibitemOpen
  \bibinfo {note} {Since the spectrum of the Fibonacci chain has zero Lebesgue
  measure~\cite {Jagannathan_RMP_2021}, there is no continuum of states, even
  for an infinite chain. We use the term \protect \textit {quasicontinuum} to
  refer to the fact that edge states in gaps much smaller than $\Delta $
  present a negligible contribution to the supercurrent and can thus be
  ignored.}\BibitemShut {Stop}%
\bibitem [{\citenamefont {Cayao}\ \emph {et~al.}(2017)\citenamefont {Cayao},
  \citenamefont {San-Jose}, \citenamefont {Black-Schaffer}, \citenamefont
  {Aguado},\ and\ \citenamefont {Prada}}]{PhysRevB.96.205425}%
  \BibitemOpen
  \bibfield  {author} {\bibinfo {author} {\bibfnamefont {J.}~\bibnamefont
  {Cayao}}, \bibinfo {author} {\bibfnamefont {P.}~\bibnamefont {San-Jose}},
  \bibinfo {author} {\bibfnamefont {A.~M.}\ \bibnamefont {Black-Schaffer}},
  \bibinfo {author} {\bibfnamefont {R.}~\bibnamefont {Aguado}},\ and\ \bibinfo
  {author} {\bibfnamefont {E.}~\bibnamefont {Prada}},\ }\bibfield  {title}
  {\bibinfo {title} {Majorana splitting from critical currents in {Josephson}
  junctions},\ }\href {https://doi.org/10.1103/PhysRevB.96.205425} {\bibfield
  {journal} {\bibinfo  {journal} {Phys. Rev. B}\ }\textbf {\bibinfo {volume}
  {96}},\ \bibinfo {pages} {205425} (\bibinfo {year} {2017})}\BibitemShut
  {NoStop}%
\bibitem [{Note2()}]{Note2}%
  \BibitemOpen
  \bibinfo {note} {We note that for finite-size \glspl {jj} there is a finite
  contribution to the current from states above the gap that is only canceled
  out after considering all the energy levels. This effect explains the values
  $I_n/I_c>1$ for finite $n$, which do not mean that the current increases over
  the maximum current that flows across the junction.}\BibitemShut {Stop}%
\bibitem [{\citenamefont {Dvir}\ \emph {et~al.}(2023)\citenamefont {Dvir},
  \citenamefont {Wang}, \citenamefont {van Loo}, \citenamefont {Liu},
  \citenamefont {Mazur}, \citenamefont {Bordin}, \citenamefont {Ten~Haaf},
  \citenamefont {Wang}, \citenamefont {van Driel}, \citenamefont {Zatelli},
  \citenamefont {Li}, \citenamefont {Malinowski}, \citenamefont {Gazibegovic},
  \citenamefont {Badawy}, \citenamefont {Bakkers}, \citenamefont {Wimmer},\
  and\ \citenamefont {Kouwenhoven}}]{Dvir2023Feb}%
  \BibitemOpen
  \bibfield  {author} {\bibinfo {author} {\bibfnamefont {T.}~\bibnamefont
  {Dvir}}, \bibinfo {author} {\bibfnamefont {G.}~\bibnamefont {Wang}}, \bibinfo
  {author} {\bibfnamefont {N.}~\bibnamefont {van Loo}}, \bibinfo {author}
  {\bibfnamefont {C.-X.}\ \bibnamefont {Liu}}, \bibinfo {author} {\bibfnamefont
  {G.~P.}\ \bibnamefont {Mazur}}, \bibinfo {author} {\bibfnamefont
  {A.}~\bibnamefont {Bordin}}, \bibinfo {author} {\bibfnamefont {S.~L.~D.}\
  \bibnamefont {Ten~Haaf}}, \bibinfo {author} {\bibfnamefont {J.-Y.}\
  \bibnamefont {Wang}}, \bibinfo {author} {\bibfnamefont {D.}~\bibnamefont {van
  Driel}}, \bibinfo {author} {\bibfnamefont {F.}~\bibnamefont {Zatelli}},
  \bibinfo {author} {\bibfnamefont {X.}~\bibnamefont {Li}}, \bibinfo {author}
  {\bibfnamefont {F.~K.}\ \bibnamefont {Malinowski}}, \bibinfo {author}
  {\bibfnamefont {S.}~\bibnamefont {Gazibegovic}}, \bibinfo {author}
  {\bibfnamefont {G.}~\bibnamefont {Badawy}}, \bibinfo {author} {\bibfnamefont
  {E.~P. A.~M.}\ \bibnamefont {Bakkers}}, \bibinfo {author} {\bibfnamefont
  {M.}~\bibnamefont {Wimmer}},\ and\ \bibinfo {author} {\bibfnamefont {L.~P.}\
  \bibnamefont {Kouwenhoven}},\ }\bibfield  {title} {\bibinfo {title}
  {{Realization of a minimal Kitaev chain in coupled quantum dots}},\ }\href
  {https://doi.org/10.1038/s41586-022-05585-1} {\bibfield  {journal} {\bibinfo
  {journal} {Nature}\ }\textbf {\bibinfo {volume} {614}},\ \bibinfo {pages}
  {445} (\bibinfo {year} {2023})}\BibitemShut {NoStop}%
\bibitem [{\citenamefont {Bordin}\ \emph {et~al.}(2024)\citenamefont {Bordin},
  \citenamefont {Li}, \citenamefont {van Driel}, \citenamefont {Wolff},
  \citenamefont {Wang}, \citenamefont {ten Haaf}, \citenamefont {Wang},
  \citenamefont {van Loo}, \citenamefont {Kouwenhoven},\ and\ \citenamefont
  {Dvir}}]{Bordin_PRL2024}%
  \BibitemOpen
  \bibfield  {author} {\bibinfo {author} {\bibfnamefont {A.}~\bibnamefont
  {Bordin}}, \bibinfo {author} {\bibfnamefont {X.}~\bibnamefont {Li}}, \bibinfo
  {author} {\bibfnamefont {D.}~\bibnamefont {van Driel}}, \bibinfo {author}
  {\bibfnamefont {J.~C.}\ \bibnamefont {Wolff}}, \bibinfo {author}
  {\bibfnamefont {Q.}~\bibnamefont {Wang}}, \bibinfo {author} {\bibfnamefont
  {S.~L.~D.}\ \bibnamefont {ten Haaf}}, \bibinfo {author} {\bibfnamefont
  {G.}~\bibnamefont {Wang}}, \bibinfo {author} {\bibfnamefont {N.}~\bibnamefont
  {van Loo}}, \bibinfo {author} {\bibfnamefont {L.~P.}\ \bibnamefont
  {Kouwenhoven}},\ and\ \bibinfo {author} {\bibfnamefont {T.}~\bibnamefont
  {Dvir}},\ }\bibfield  {title} {\bibinfo {title} {Crossed {Andreev} reflection
  and elastic cotunneling in three quantum dots coupled by superconductors},\
  }\href {https://doi.org/10.1103/PhysRevLett.132.056602} {\bibfield  {journal}
  {\bibinfo  {journal} {Phys. Rev. Lett.}\ }\textbf {\bibinfo {volume} {132}},\
  \bibinfo {pages} {056602} (\bibinfo {year} {2024})}\BibitemShut {NoStop}%
\bibitem [{\citenamefont {Zatelli}\ \emph {et~al.}(2024)\citenamefont
  {Zatelli}, \citenamefont {van Driel}, \citenamefont {Xu}, \citenamefont
  {Wang}, \citenamefont {Liu}, \citenamefont {Bordin}, \citenamefont {Roovers},
  \citenamefont {Mazur}, \citenamefont {van Loo}, \citenamefont {Wolff},
  \citenamefont {Bozkurt}, \citenamefont {Badawy}, \citenamefont {Gazibegovic},
  \citenamefont {Bakkers}, \citenamefont {Wimmer}, \citenamefont
  {Kouwenhoven},\ and\ \citenamefont {Dvir}}]{Zatelli2024Sep}%
  \BibitemOpen
  \bibfield  {author} {\bibinfo {author} {\bibfnamefont {F.}~\bibnamefont
  {Zatelli}}, \bibinfo {author} {\bibfnamefont {D.}~\bibnamefont {van Driel}},
  \bibinfo {author} {\bibfnamefont {D.}~\bibnamefont {Xu}}, \bibinfo {author}
  {\bibfnamefont {G.}~\bibnamefont {Wang}}, \bibinfo {author} {\bibfnamefont
  {C.-X.}\ \bibnamefont {Liu}}, \bibinfo {author} {\bibfnamefont
  {A.}~\bibnamefont {Bordin}}, \bibinfo {author} {\bibfnamefont
  {B.}~\bibnamefont {Roovers}}, \bibinfo {author} {\bibfnamefont {G.~P.}\
  \bibnamefont {Mazur}}, \bibinfo {author} {\bibfnamefont {N.}~\bibnamefont
  {van Loo}}, \bibinfo {author} {\bibfnamefont {J.~C.}\ \bibnamefont {Wolff}},
  \bibinfo {author} {\bibfnamefont {A.~M.}\ \bibnamefont {Bozkurt}}, \bibinfo
  {author} {\bibfnamefont {G.}~\bibnamefont {Badawy}}, \bibinfo {author}
  {\bibfnamefont {S.}~\bibnamefont {Gazibegovic}}, \bibinfo {author}
  {\bibfnamefont {E.~P. A.~M.}\ \bibnamefont {Bakkers}}, \bibinfo {author}
  {\bibfnamefont {M.}~\bibnamefont {Wimmer}}, \bibinfo {author} {\bibfnamefont
  {L.~P.}\ \bibnamefont {Kouwenhoven}},\ and\ \bibinfo {author} {\bibfnamefont
  {T.}~\bibnamefont {Dvir}},\ }\bibfield  {title} {\bibinfo {title} {{Robust
  poor man{'}s Majorana zero modes using Yu-Shiba-Rusinov states}},\ }\href
  {https://doi.org/10.1038/s41467-024-52066-2} {\bibfield  {journal} {\bibinfo
  {journal} {Nat. Commun.}\ }\textbf {\bibinfo {volume} {15}},\ \bibinfo
  {pages} {1} (\bibinfo {year} {2024})}\BibitemShut {NoStop}%
\bibitem [{\citenamefont {Tombros}\ \emph {et~al.}(2011)\citenamefont
  {Tombros}, \citenamefont {Veligura}, \citenamefont {Junesch}, \citenamefont
  {Jasper van~den Berg}, \citenamefont {Zomer}, \citenamefont {Wojtaszek},
  \citenamefont {Vera~Marun}, \citenamefont {Jonkman},\ and\ \citenamefont {van
  Wees}}]{Tombros_JApplPhys2011May}%
  \BibitemOpen
  \bibfield  {author} {\bibinfo {author} {\bibfnamefont {N.}~\bibnamefont
  {Tombros}}, \bibinfo {author} {\bibfnamefont {A.}~\bibnamefont {Veligura}},
  \bibinfo {author} {\bibfnamefont {J.}~\bibnamefont {Junesch}}, \bibinfo
  {author} {\bibfnamefont {J.}~\bibnamefont {Jasper van~den Berg}}, \bibinfo
  {author} {\bibfnamefont {P.~J.}\ \bibnamefont {Zomer}}, \bibinfo {author}
  {\bibfnamefont {M.}~\bibnamefont {Wojtaszek}}, \bibinfo {author}
  {\bibfnamefont {I.~J.}\ \bibnamefont {Vera~Marun}}, \bibinfo {author}
  {\bibfnamefont {H.~T.}\ \bibnamefont {Jonkman}},\ and\ \bibinfo {author}
  {\bibfnamefont {B.~J.}\ \bibnamefont {van Wees}},\ }\bibfield  {title}
  {\bibinfo {title} {Large yield production of high mobility freely suspended
  graphene electronic devices on a polydimethylglutarimide based organic
  polymer},\ }\href {https://doi.org/10.1063/1.3579997} {\bibfield  {journal}
  {\bibinfo  {journal} {Journal of Applied Physics}\ }\textbf {\bibinfo
  {volume} {109}},\ \bibinfo {pages} {093702} (\bibinfo {year}
  {2011})}\BibitemShut {NoStop}%
\bibitem [{\citenamefont {Buscema}\ \emph {et~al.}(2014)\citenamefont
  {Buscema}, \citenamefont {Groenendijk}, \citenamefont {Steele}, \citenamefont
  {van~der Zant},\ and\ \citenamefont
  {Castellanos-Gomez}}]{Buscema_Nature2014}%
  \BibitemOpen
  \bibfield  {author} {\bibinfo {author} {\bibfnamefont {M.}~\bibnamefont
  {Buscema}}, \bibinfo {author} {\bibfnamefont {D.~J.}\ \bibnamefont
  {Groenendijk}}, \bibinfo {author} {\bibfnamefont {G.~A.}\ \bibnamefont
  {Steele}}, \bibinfo {author} {\bibfnamefont {H.~S.~J.}\ \bibnamefont {van~der
  Zant}},\ and\ \bibinfo {author} {\bibfnamefont {A.}~\bibnamefont
  {Castellanos-Gomez}},\ }\bibfield  {title} {\bibinfo {title} {{Photovoltaic
  effect in few-layer black phosphorus PN junctions defined by local
  electrostatic gating}},\ }\href {https://doi.org/10.1038/ncomms5651}
  {\bibfield  {journal} {\bibinfo  {journal} {Nat. Commun.}\ }\textbf {\bibinfo
  {volume} {5}},\ \bibinfo {pages} {1} (\bibinfo {year} {2014})}\BibitemShut
  {NoStop}%
\bibitem [{\citenamefont {Maurand}\ \emph {et~al.}(2014)\citenamefont
  {Maurand}, \citenamefont {Rickhaus}, \citenamefont {Makk}, \citenamefont
  {Hess}, \citenamefont
  {T{\ifmmode\acute{o}\else\'{o}\fi}v{\ifmmode\acute{a}\else\'{a}\fi}ri},
  \citenamefont {Handschin}, \citenamefont {Weiss},\ and\ \citenamefont
  {Sch{\ifmmode\ddot{o}\else\"{o}\fi}nenberger}}]{Maurand_JCarbon2014}%
  \BibitemOpen
  \bibfield  {author} {\bibinfo {author} {\bibfnamefont {R.}~\bibnamefont
  {Maurand}}, \bibinfo {author} {\bibfnamefont {P.}~\bibnamefont {Rickhaus}},
  \bibinfo {author} {\bibfnamefont {P.}~\bibnamefont {Makk}}, \bibinfo {author}
  {\bibfnamefont {S.}~\bibnamefont {Hess}}, \bibinfo {author} {\bibfnamefont
  {E.}~\bibnamefont
  {T{\ifmmode\acute{o}\else\'{o}\fi}v{\ifmmode\acute{a}\else\'{a}\fi}ri}},
  \bibinfo {author} {\bibfnamefont {C.}~\bibnamefont {Handschin}}, \bibinfo
  {author} {\bibfnamefont {M.}~\bibnamefont {Weiss}},\ and\ \bibinfo {author}
  {\bibfnamefont {C.}~\bibnamefont
  {Sch{\ifmmode\ddot{o}\else\"{o}\fi}nenberger}},\ }\bibfield  {title}
  {\bibinfo {title} {{Fabrication of ballistic suspended graphene with
  local-gating}},\ }\href {https://doi.org/10.1016/j.carbon.2014.07.088}
  {\bibfield  {journal} {\bibinfo  {journal} {Carbon}\ }\textbf {\bibinfo
  {volume} {79}},\ \bibinfo {pages} {486} (\bibinfo {year} {2014})}\BibitemShut
  {NoStop}%
\bibitem [{\citenamefont {Cuevas}\ and\ \citenamefont
  {Scheer}(2017)}]{Cuevas_WS2017}%
  \BibitemOpen
  \bibfield  {author} {\bibinfo {author} {\bibfnamefont {J.~C.}\ \bibnamefont
  {Cuevas}}\ and\ \bibinfo {author} {\bibfnamefont {E.}~\bibnamefont
  {Scheer}},\ }\href {https://doi.org/10.1142/10598} {\emph {\bibinfo {title}
  {{Molecular Electronics {$\vert$} World Scientific Series in Nanoscience and
  Nanotechnology}}}},\ Vol.~\bibinfo {volume} {15}\ (\bibinfo  {publisher}
  {World Scientific Publishing Company},\ \bibinfo {address} {Singapore},\
  \bibinfo {year} {2017})\BibitemShut {NoStop}%
\bibitem [{\citenamefont {de~Gennes}(1966)}]{deGennes_WAB_1966}%
  \BibitemOpen
  \bibfield  {author} {\bibinfo {author} {\bibfnamefont {P.-G.}\ \bibnamefont
  {de~Gennes}},\ }\href@noop {} {\emph {\bibinfo {title} {{Superconductivity of
  Metals and Alloys}}}}\ (\bibinfo  {publisher} {W.A. Benjamin},\ \bibinfo
  {year} {1966})\BibitemShut {NoStop}%
\bibitem [{\citenamefont {Rai}\ \emph {et~al.}(2019)\citenamefont {Rai},
  \citenamefont {Haas},\ and\ \citenamefont {Jagannathan}}]{Rai_PRB2019}%
  \BibitemOpen
  \bibfield  {author} {\bibinfo {author} {\bibfnamefont {G.}~\bibnamefont
  {Rai}}, \bibinfo {author} {\bibfnamefont {S.}~\bibnamefont {Haas}},\ and\
  \bibinfo {author} {\bibfnamefont {A.}~\bibnamefont {Jagannathan}},\
  }\bibfield  {title} {\bibinfo {title} {{Proximity effect in a
  superconductor-quasicrystal hybrid ring}},\ }\href
  {https://doi.org/10.1103/PhysRevB.100.165121} {\bibfield  {journal} {\bibinfo
   {journal} {Phys. Rev. B}\ }\textbf {\bibinfo {volume} {100}},\ \bibinfo
  {pages} {165121} (\bibinfo {year} {2019})}\BibitemShut {NoStop}%
\end{thebibliography}


%%%%%%%%%%%%%%%%%%%%%%%%%%%%%%%%%%%%%%%%%%%%%%%%%%%%%%%%%%%
%%%%%%%%%%%%%%%%%%%%%%%%%%%%%%%%%%%%%%%%%%%%%%%%%%%%%%%%%%%

%%%%%%%%%%%%%%%%%%%%%%%%%%%
%%%%%%%%%%%%%%%%%%%%%%%%%%%
%%%%%%%%%%%%%%%%%%%%%%%%%%%
%%%%%%%%%%%%%%%%%%%%%%%%%%%

\clearpage

\onecolumngrid

\setcounter{equation}{0}
\renewcommand{\theequation}{S\arabic{equation}}
\crefname{equation}{Eq.}{Eqs.}
\Crefname{equation}{Equation}{Equations}

\setcounter{figure}{0}
\renewcommand{\thefigure}{S\,\arabic{figure}}
\crefname{figure}{Fig.}{Figs.}
\Crefname{figure}{Figure}{Figures}

\section{Supplemental Material for ``The Josephson effect in Fibonacci superconductors''}

In this supplementary material, we provide additional details about (1) how we define the Fibonacci sequence and the approximants; (2) the impact of the quasiperiodic potential on the junction transmission; (3) how superconducting correlations develop inside the Fibonacci gaps; (4) how Fibonacci-Andreev bound states become the dominant contribution to the supercurrent; (5) the stability of the Fibonacci gaps; and (6) the self-consistency of the superconducting order parameter. 

\subsection{The Fibonacci sequence and approximants}

The Fibonacci model is a useful representation of one-dimensional quasiperiodic chains. In the Fibonacci tight-binding chain considered here, two types of atoms, $A$ and $B$, are arranged according to the Fibonacci sequence. 
One way to generate the Fibonacci sequence is using the substitution rule, where each element of a new sequence is created substituting the elements of the previous sequence according to the rule
$$
B\to A, \ A \to AB .
$$ 
The smallest chains with one and two sites are assumed to be $A$ and $AB$. Following the substitution rule, the next chain has three sites, $ABA$, while the fifth Fibonacci chain has $8$ sites arranged as $ABAABABA$. 

We label the Fibonacci numbers as $F_0, F_1, F_2, F_2, \dots = 1, 2, 3, 5, \dots$, where $F_n = F_{n-1} + F_{n-2}$, with $F_0=1$ and $F_1=2$ by definition. 
Then, after $n$ generations the number of sites is $F_n$ and the chain is composed of $F_{n-1} $ $A$ sites and $F_{n-2}$ $B$ sites arranged in a Fibonacci sequence. 

In the main text, the distinction between $A$ and $B$ atoms is their onsite energy $\varepsilon_{a,b}$. 
A chain with $F_n$ sites thus has $F_{n-1}$ sites with onsite potential $\varepsilon_a$ and $F_{n-2}$ sites with energy $\varepsilon_b$. 
Imposing that the mean onsite potential in the chain is zero (charge neutrality), our choice for the ratio between the onsite potentials is $\varepsilon_b / \varepsilon_a = - F_{n-2}/F_{n-1}$. 
Finally, a chain with $F_n$ sites still presents different arrangements of the Fibonacci sequence depending on the phason angle $\theta$. The approximant $S_n$ indicates the set of Fibonacci chains with the same number of sites $F_n$ but different phason angles. 

\subsection{Normal-state transmission}

The transmission between two Fibonacci chains in the normal state strongly depends on the onsite potentials at the junction. 
In the main text, we demonstrated that the effective transparency of the \gls{abs} and the overall critical current of the junction are strongly influenced by the phason angle, which determines the values of the onsite energies at the sites forming the junction interface. 
In this section, we explore in more detail the impact of the quasicrystalline potential texture on the transmission coefficient. For simplicity, we only consider Fibonacci chains in the normal state and doped at the Fermi level. 

We consider a symmetric junction formed by two finite Fibonacci chains with approximant $S_n$, labeled $L$ and $R$, coupled by a tunnel potential $V_{LR}=V_{RL}=t_0$. We compute the transmission following the usual expression from Keldysh formalism~\cite{Cuevas_WS2017},
\begin{equation} 
    T(E) = \Tr \{ \Gamma_L(E) G^r_{RL}(E) \Gamma_R(E) [G^r_{RL}(E)]^\dagger \}, 
\end{equation}
where we have defined the couplings $\Gamma_{\alpha}=\mathrm{Im} \{ t_0 g_{\bar{\alpha}\bar{\alpha}} t_0 \}$, with $\alpha=L,R$ and $\bar\alpha$ the opposite, and the coupled retarded and advanced Green functions using Dyson equation, $G_{\alpha\bar{\alpha}}^{r,a} = g_{\alpha\alpha}^{r,a} t_0 G^{r,a}_{\bar{\alpha}\bar{\alpha}}$, with $G^{r,a}_{\alpha\alpha} = \left[ (g^{r,a}_{\alpha\alpha})^{-1} - t_0 g^{r,a}_{\bar{\alpha}\bar{\alpha}} t_0 \right]^{-1}$. Here, $g^{r,a}_{\alpha\alpha}$ are the retarded and advanced local Green functions at the edges of the decoupled Fibonacci chains ($g^{r,a}_{LR}=g^{r,a}_{RL}=0$). 

\begin{figure}[t]
    \includegraphics[width=\linewidth]{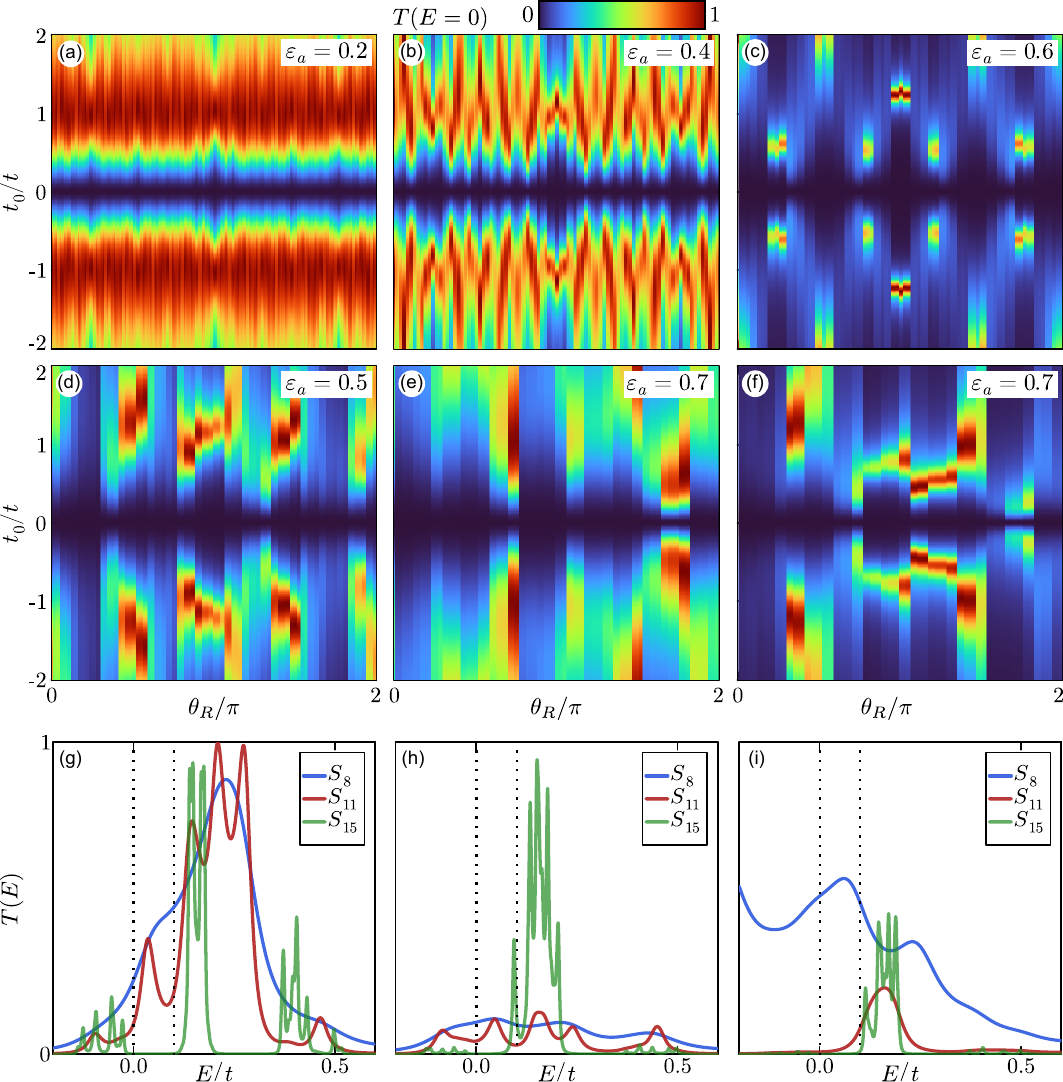}
	\caption{Transmission of the junction in the normal state. 
    (a-f) Zero-frequency transmission $T(E=0)$ as a function of right phason angle $\theta_R$ and hopping between chains $t_0$. (a-c) Symmetric case with $\theta_L = \theta_R$ and $\varepsilon_a/t=0.2$ (a), $0.4$ (b) and $0.6$ (c). (d-f) Asymmetric cases with $\theta_L=0.5\pi$ (d,e) and $\theta_L=0.9\pi$ (f). In all cases, $\varepsilon_b / \sum_i\varepsilon_i=0$. (g-i) Frequency dependence of the transmission $T(E)$ at different phason angles for the parameters used in \cref{fig2} of the main text: $t_0=0.6t$, $\varepsilon_a=0.7t$, $\Delta=0.1t$, $\theta_L = 0.9\pi$, and $\theta_R = 0.45\pi$ (g), $\theta_R = 0.2\pi$ (h) and $\theta_R=1.8\pi$ (i). Each line is computed for a different Fibonacci approximant. Dashed vertical lines mark $E=0$ and $E=\Delta=0.1t$, as in the main text. 
 }
	\label{fig:transmission}
\end{figure}

The uncoupled Green functions $g^{r,a}_{\alpha\alpha}$ are computed iteratively. For example, for chain $L$ with sites going from $1$ to $F_n$, we have 
\begin{subequations}
    \begin{align}
	g_{11} (E) ={}&  \left( E - \varepsilon_a \right)^{-1}, \\
	g_{LL}\equiv g_{F_{n}F_{n}} (E) ={}& \left( E - \varepsilon_{F_{n}} - t^2 g_{F_{n}-1,F_{n}-1} \right)^{-1}, 
\end{align}
\end{subequations}
with $t$ the hopping amplitude. The edge Green function $g_{RR}$ is obtained following an equivalent procedure: Starting from the rightmost edge, 
\begin{equation}
    g_{F_{n}F_{n}} (E) = \left( E - \varepsilon_{F_{n}} \right)^{-1}, 
\end{equation}
and iterating until reaching the leftmost edge, $g_{RR}= g_{11}$. 

We compute the transmission as a function of $\theta_R$ and $t_0$ in \cref{fig:transmission}(a-f). We consider two complementary scenarios: (1) both chains have the same phason angle, exhibiting the usual symmetry around $\theta = \pi$; and (2) the phason angle in the right chain $\theta_R$ is changed while keeping the one on the left fixed at $\theta_L$, breaking the symmetry around $\theta_R = \pi$. 

In the symmetric case, \cref{fig:transmission}(a-c) with $\theta_L = \theta_R$, the transmission as a function of $t_0$ becomes progressively narrower as we increase $\varepsilon_a$. This is the result of more Fibonacci gaps opening in the spectrum as $\varepsilon_a$ increases. At $\varepsilon_a=0.6$, \cref{fig:transmission}(c), the transmission is still high ($T(E=0) \sim 1$) on resonance, that is, when the levels between Fibonacci gaps from the left and right side align around zero energy. Since both halves of the junction satisfy the resonance condition at the same phason angle, the resulting resonance peak is relatively narrow.
By contrast, in the asymmetric case, \cref{fig:transmission}(d-f) with $\theta_L \neq \theta_R$, the dependence of the transmission on $t_0$ varies significantly with the choice of the fixed $\theta_L$, featuring more resonant behavior than in the symmetric case. 

We also compute the transmission as a function of the energy in \cref{fig:transmission}(g-i), using the same parameters as in \cref{fig4}(a) and \cref{fig4}(c) of the main text ($S_{11}$ for red lines). The first case (\textcircled{\raisebox{-0.5pt}{1}}) corresponds to a highly transmitting junction dominated by \glspl{abs} [\cref{fig4}(a)] and the other (\textcircled{\raisebox{-0.5pt}{3}}) to a tunnel junction with a current entirely given by \glspl{fabs} contributions. 
We can thus attribute the high effective transmission of \gls{abs} in case \textcircled{\raisebox{-0.5pt}{1}} in the main text to the presence of transmission peaks close to the Fermi level $E=0$, smaller than $\Delta$, panel (g), while in the case of \textcircled{\raisebox{-0.5pt}{3}} the peaks are further away in energies of the order of $\Delta$, panel (i).

\begin{figure}[b]
    \includegraphics[width=\linewidth]{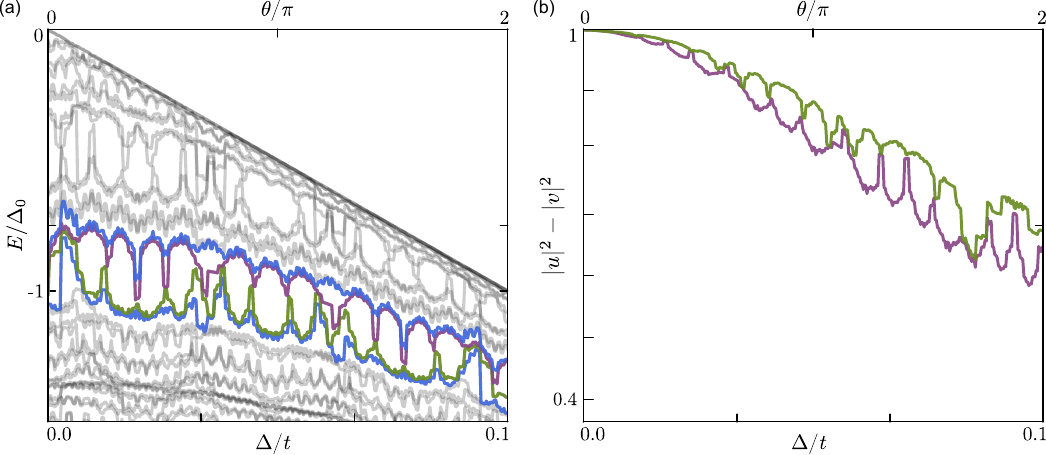}
	\caption{(a) Energy levels of the $S_{11}$ approximant as a function of the superconducting gap and the phason angle. One of the low-energy Fibonacci gaps is highlighted in blue, with its edge modes in purple and green. (b) Charge of the Fibonacci edge modes as a function of $\theta$ and $\Delta$. We fix $\Delta_0=0.1t$, equal to the gap amplitude used in the main text.
    }
	\label{fig:Charge}
\end{figure}

\subsection{Electron-hole correlations in the Fibonacci gaps}

In the main text, we argued that \glspl{fabs} are Fibonacci quasicrystal states that behave as Andreev bound states. This behavior is made possible through the proximity effect, where superconducting correlations extend beyond the superconducting gap to higher energies, forming the \glspl{fabs}. In this section, we support this picture by analyzing the electron-hole content of the Fibonacci edge states as a function of the pairing potential, \cref{fig:Charge}. To do so, we examine the charge distribution of the eigenstates in one of the low-energy gaps of the $S_{11}$ approximant. 

First, we plot the bands for $S_{11}$ in \cref{fig:Charge}(a) as a function of both the phason angle $\theta$ and the superconducting pairing potential $\Delta$. 
Since we are varying $\Delta$ starting from $\Delta=0$, we scale the energy to a fixed $\Delta_0=0.1t$, the same gap value used in the main text. 
For a given stable Fibonacci gap at energy above $\Delta_0$, we highlight the topological states using purple and green lines. The gap edge is marked by blue lines. The edge states do not disappear when we turn on superconductivity, neither is the stable gap closed, although the spectrum is shifted toward higher energies due to the presence of the superconducting gap. 

The effective charge of the topological Fibonacci states is shown in \cref{fig:Charge}(b), indicating its electron-hole content. The wavefunction corresponding to the energy level $\epsilon_n$, in the Nambu basis, can be written as $\bra{\psi_n} = (u_n^1, v_n^1, ..., u_n^N, v_n^N)$, since it has an electron-hole degree of freedom [$(u,v)$] in each of the $N$ sites of the chain. We therefore define the state electronic content as $|u_n|^2=\sum_i|u_n^i|^2$, and similarly for the hole content $|v_n|^2$. The state charge is thus $|u_n|^2-|v_n|^2$.
As expected, the states are purely electronic ($|v|^2=0$) for $\Delta=0$, see green and purple lines in \cref{fig:Charge}(b). 
Even though these states appear at energies well above the gap, the fully electronic states at $\Delta=0$ develop a finite hole component at finite $\Delta$.

\begin{figure}[t]
	\centering
    \includegraphics[width=1.0\linewidth]{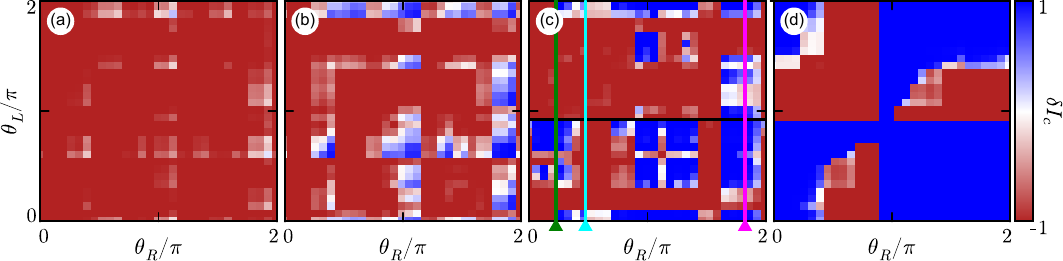}
	\caption{Difference between \gls{abs} and \gls{fabs} critical currents $\delta I_c$ as a function of the phason angles $\theta_L$ and $\theta_R$. Blue regions indicate \glspl{fabs} dominance ($\delta I_c>0$). (a) $\varepsilon_a/t=0.3$, (b) $\varepsilon_a/t=0.5$, (c) $\varepsilon_a/t=0.7$, and (d) $\varepsilon_a/t=1.0$. 
    In panel (c), the intersection between the horizontal black line ($\theta_L/\pi=0.9$) and the vertical colored lines correspond to the Josephson currents shown in \cref{fig4}(a-c) of the main text: cyan at $\theta_R/\pi = 0.45$ for \cref{fig4}(a), green at $\theta_R/\pi = 0.2$for \cref{fig4}(b), and magenta at $\theta_R/\pi = 1.8$ for \cref{fig4}(c).
 }
	\label{fig:FABSRegions}
\end{figure}

\subsection{Supercurrent dominated by Fibonacci-Andreev Bound States}\label{sec:dominance}

One of our main results is that \glspl{fabs} can dominate the supercurrent when the phason angles favor the emergence of topological states at the junction interface. In this section, we explore in more detail the emergence of a supercurrent dominated by \glspl{fabs}. To do so, we define 
\begin{equation}
    \delta I_c = \frac{I_c^\text{FABS} - I_c^\text{ABS}} {I_c} ,
\end{equation}
which measures the difference between the critical current contributions from \glspl{abs} and \glspl{fabs}, in units of $I_c = \max_\phi \left\{ \sum_n\frac{dE_n}{d\phi} \right\}=I_c^\text{FABS} + I_c^\text{ABS}$. 
Specifically, we define 
\begin{equation}
    I_c^{\mathrm{ABS}} =  \left( \frac{\md E_1}{\md \phi} \right)_{\phi=\phi_c} , \quad \text{and}  \quad 
    I_c^{\mathrm{FABS}} = \left( \sum_{n=2}^\infty \frac{\md E_n}{\md \phi} \right)_{\phi=\phi_c} ,
\end{equation}
where $\phi_{c}$ is determined by the condition $I(\phi_{c})=I_{c}$, and we label $E_1<0$ the highest of all $E_n<0$ states contributing to the current. 

We analyze $\delta I_c$ in \cref{fig:FABSRegions} mapping it as a function of phason angles $\theta_L$ and $\theta_R$ for a fixed, intermediate transparency $t_0 = 0.6t$, the same as in the main text. 
For each panel we increase the strength of the Fibonacci electrostatic potential, with \cref{fig:FABSRegions}(c) corresponding to the value of $\varepsilon_a/t=0.7$ used in the main text. 
As our analysis of the normal-state transmission revealed, different regions of high and low transmission emerge depending on the combinations of $\theta_L$ and $\theta_R$. 
As the quasicrystal parameter $\varepsilon_a$ increases, the transmission becomes more resonant suppressing the \gls{abs} contribution and enhancing the \glspl{fabs} one (blue regions). The \glspl{fabs} dominance becomes more pronounced as the Fibonacci gaps grow in size. At the same time, since the junction approaches the tunnel regime, the total critical current $I_c$ decreases. This is consistent with the result in the main text that the Fibonacci-dominated current is always tunnel (the current-phase relation is sinusoidal).

\begin{figure}[ht]
    \includegraphics[width=1.0\linewidth]{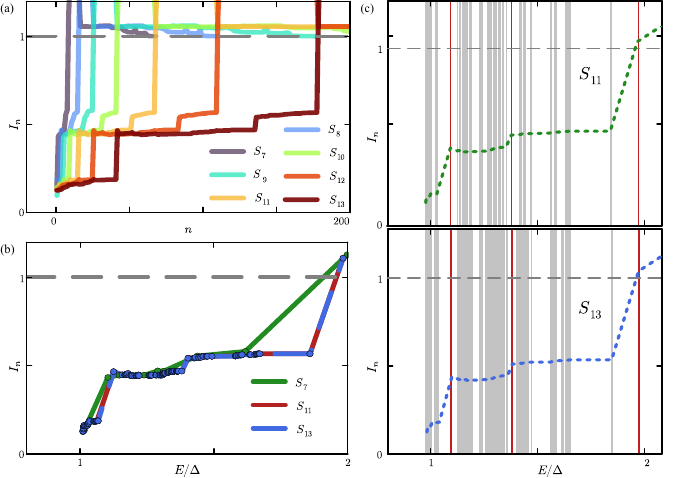}
	\caption{(a) Accumulated critical current $I_n$ as a function of the eigenvalue label $n$ for different approximants. Gray dashed line indicates the total critical current. (b) Same as (a) as a function of the energy $E$. The results for $S_{11}$ used in the main text have reached convergence; even $S_{7}$ already has the main contributions. (c) $I_n$ vs $E$ curve for $S_{13}$ (blue) on top of the spectrum of the 13th approximant (gray and red vertical lines). The red lines are the \glspl{fabs} localized close to the interface, that contribute  to the supercurrent.
 }
	\label{fig:Cumulants}
\end{figure}

\subsection{Stability of \glspl{fabs} for larger approximants}

We argued in the main text that \glspl{fabs} formed at Fibonacci gaps can carry the supercurrent. However, the number of gaps in the spectrum of a Fibonacci chain depends on its size, or approximant. 
A Fibonacci gap appearing for two consecutive approximants is considered stable, i.e., persisting in the thermodynamic limit. 
We thus ensure here that the approximant used in the main text is such that the \glspl{fabs} contributions to the supercurrent are stable. 

We plot in \cref{fig:Cumulants} the contribution to the supercurrent from different states, represented by the quantity $I_n$ defined in \cref{eq:In} in the main text. We compute this quantity for different approximants $S_n$ to check the stability of the current carried by different states. 

\Cref{fig:Cumulants}(a) shows $I_n$ as a function of $n$ computed for several approximants; from short ones like $S_{7}$ where each chain has $34$ sites, to longer ones like $S_{13}$ with $610$ sites per chain. For all cases, the initial value at $n=1$, representing the contribution to the critical current from the \gls{abs}, is roughly the same: $I_1\sim0.25I_c$. The next contributions, marked by jumps in $I_n$ take place for different states depending on the approximant, but they all have similar values. The different state number is the result of the chain having different size for each approximant. To better visualize this, we plot $I_n$ as a function of the state energy in \cref{fig:Cumulants}(b). We see now that the jumps take place at similar energies, so the gaps hosting the \glspl{fabs} contributing to the current appear at the same energy for different approximants. 

These results confirm that the approximant $S_{11}$ ($233$ sites) used in the main text is sufficient for convergence. Notably, even shorter approximants like $S_7$ exhibit qualitatively similar behavior. This fast convergence is due to the largest \gls{fabs} contributions occurring for the largest stable Fibonacci gaps, which emerge early in the sequence of approximants. 
This is illustrated in \cref{fig:Cumulants}(c), where we compare the spectrum (gray lines) for the $S_{11}$ (top) and $S_{13}$ (bottom) approximants with the current contribution from each state (colored dashed lines). The Fibonacci gaps are stable since they appear for both approximants. The jumps in $I_n$ clearly correspond to the states inside them (red lines). 

\subsection{Currents with self-consistent order parameter}
In this section we verify that the results presented in the main text remain qualitatively unchanged when the superconducting order parameter is treated self-consistently.

The system is described by the same Hamiltonian as in the main text, \cref{eq:fullJJ}, 
\begin{equation}
    H = H_L + H_R + H_T,
    \label{eq:fullJJ2}
\end{equation}
However, we now include a site-dependent mean-field repulsion $U_i$ and pair amplitude $\Delta_i$, together with the paring potential $V$~\cite{deGennes_WAB_1966,Rai_PRB2019},
\begin{equation}
 \label{eq:HLR2}
	H_{\alpha} = \sum_{i=1}^{{N}_{\alpha}} [(\varepsilon_i - \mu + U_i) c_{i\sigma}^\dagger c_{i\sigma} 
	- t (c_{i+1\sigma}^\dagger c_{i\sigma} + \mathrm{H.c.})] 
	+ V \sum_{i=1}^{{N}_{\alpha}} \Delta^i c_{i\sigma}^\dagger c_{i\bar{\sigma}}^\dagger + \mathrm{H.c.}~,
\end{equation}

\begin{figure}[t]	
	\includegraphics[width=1.\linewidth]{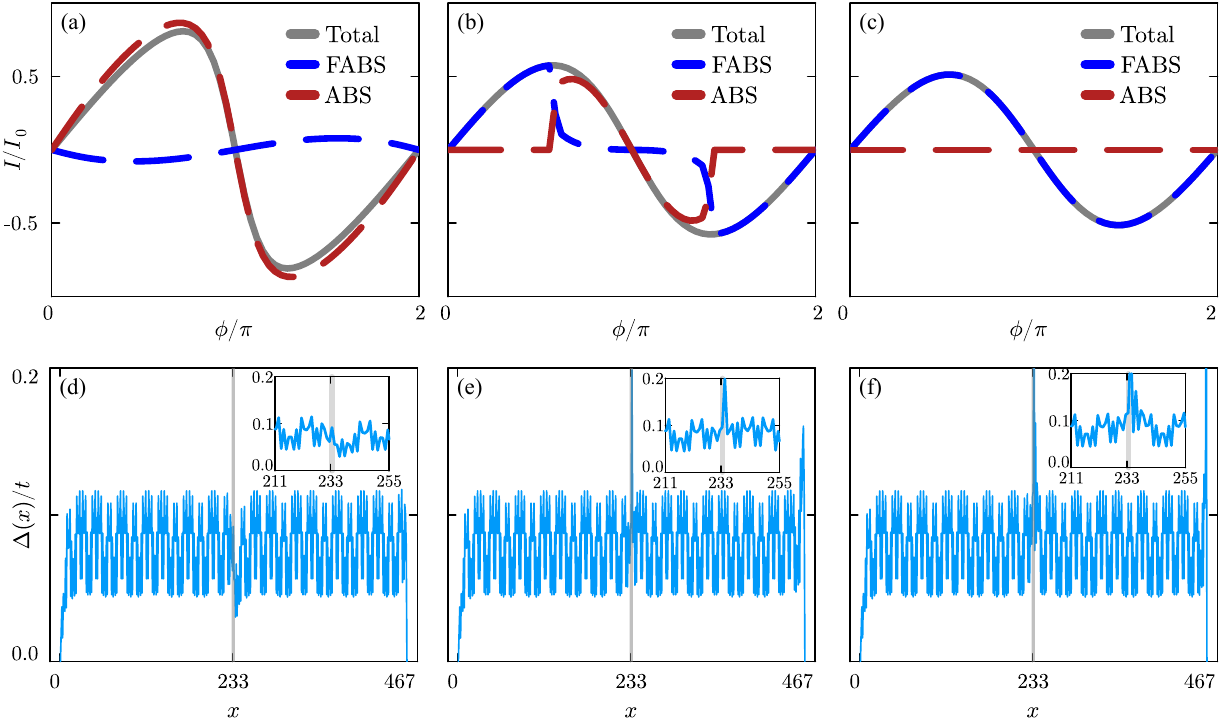}
	\caption{Self-consistent supercurrent. 
    (a-c) Current-phase relation using a self-consistent pair potential for cases \textcircled{\raisebox{-0.5pt}{1}} (a), \textcircled{\raisebox{-0.5pt}{2}} (b), and \textcircled{\raisebox{-0.5pt}{3}} (c) of the main text (total current in gray, \gls{fabs} and \gls{abs} contributions in, respectively, blue and red dashed lines). 
    (d-f) Self-consistent $\Delta(x)$ for the cases above. The insets zoom around the interface, marked by the gray vertical line. We used $V=1.6t$ and mean $\Delta(x)$ is $0.07t$. 
 }
	\label{fig:scCurrent}
\end{figure}

\begin{figure}[ht]	
	\includegraphics[width=\linewidth]{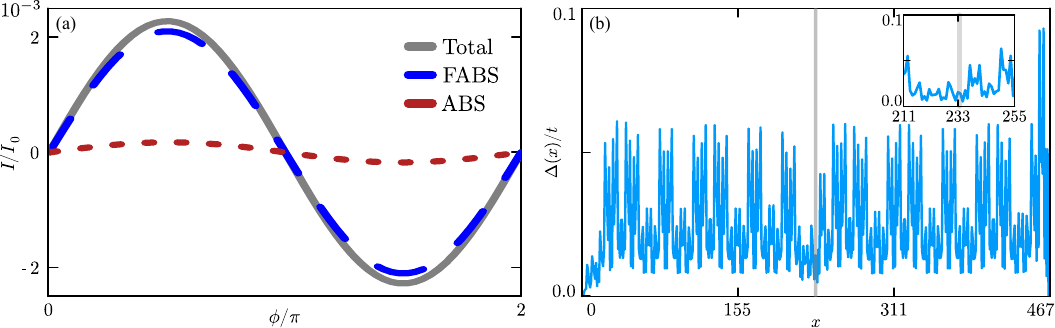}
	\caption{(a) Self-consistent supercurrent for case \textcircled{\raisebox{-0.75pt}{3}} at smaller $V=0.7$, still showing \gls{fabs} dominance. (b) self-consistent pairing potential for a mean order parameter of $0.017t$. 
 }
	\label{fig:smallerDeltaCurrent}
\end{figure}

The pair amplitude is determined self-consistently by solving the eigenvalue problem $H\ket{\psi_n}=E_n\ket{\psi_n}$. At zero temperature,
\begin{equation}
    \Delta^i = V \sum_{E_n<0} v_n^{i*} u_n^i \,
\end{equation}
where $u_{n}^i$ ($v_n^i$) is the electron (hole) component of $\ket{\psi_n}$ at site $i$. 
Similarly, the onsite repulsion is also determined self-consistently as 
\begin{equation}
    U_i = -V \left( \sum_n |u_{n}^{i}|^2 + \sum_n |v_{n}^{i}|^2 \right).
\end{equation}

Both $\Delta_i$ and $U_i$ result from the mean-field approximation of a Hubbard model for superconductivity. The phase difference between the superconductors is included in the tunneling Hamiltonian,
\begin{equation}
    H_T = -t_0 e^{i(\phi_L - \phi_R)} c_{N_L+1}^\dagger c_{N_L} + \mathrm{H.c.}
\end{equation}

Starting from a homogeneous initial profile for $\Delta(x)$ with amplitude $0.1t$ and using $V = 1.6$, we obtain convergence after 85 iterations. 
We plot the self-consistent current and the spatial profile of the converged $\Delta(x)$ in \cref{fig:scCurrent} for the cases labeled in the main text as \textcircled{\raisebox{-0.5pt}{1}}, \textcircled{\raisebox{-0.5pt}{2}} and \textcircled{\raisebox{-0.5pt}{3}}. 
In all cases, the final mean $\Delta(x)$ is approximately $0.07t$, resulting in an order parameter $V\Delta = 0.11 t$, consistent with the one shown in the main text. The profile itself shows quasiperiodic oscillations, consistent with previous literature~\cite{Rai_PRB2019,Wang_arxiv_2024}.

The self-consistent calculations for the supercurrent show behavior in line with the non-self-consistent case shown in the main text. For $\theta_R/\pi = 0.45$, case \textcircled{\raisebox{-0.5pt}{1}} in \cref{fig:scCurrent}(a), the critical current is dominated by \gls{abs} with $I_c \sim 0.75 I_0$, slightly lower than in the main text. At $\theta_R/\pi = 0.2$, case \textcircled{\raisebox{-0.5pt}{2}} in \cref{fig:scCurrent}(b), a mixed contribution gives $I_c \sim 0.5 I_0$, closely matching the non-self-consistent result. For $\theta_R/\pi = 1.8$, case \textcircled{\raisebox{-0.5pt}{3}} in \cref{fig:scCurrent}(c), the current is dominated by \gls{fabs} and also yields $I_c \sim 0.5 I_0$, consistent with the main text and showing a slight increase but no qualitative change. We also repeat the calculation for $\theta_R/\pi = 1.8$ with $V/t = 0.7$, resulting in a mean $V\Delta(x) \approx 0.017t$. The results shown in \cref{fig:smallerDeltaCurrent} confirm that reducing the order parameter does not change our results. 

\end{document}